\title{Community Enforcement with Endogenous Records}
\author{Harry PEI\footnote{Northwestern University. Email: harrydp@northwestern.edu. I thank Daron Acemoglu, S. Nageeb Ali, 
V Bhaskar,
Costas Cavounidis, Drew Fudenberg, Kevin He,
Johannes H\"{o}rner,  
Michihiro Kandori, David Levine, Alessandro Lizzeri, Daniel Luo,
Stephen Morris, Wojciech Olszewski,
Daisuke Oyama, Larry Samuelson,
Satoru Takahashi, Caroline Thomas,
Juuso V\"{a}lim\"{a}ki, and Alexander Wolitzky for helpful comments. I thank the NSF grants SES-1947021 and SES-2337566 for financial support.}}
\date{\today}
\documentclass[11pt]{article}

\usepackage{amsmath}
\usepackage{graphicx}
\usepackage{amsfonts}
\usepackage{amsthm}
\usepackage{setspace}
\usepackage{tikz}
\usepackage{amssymb}
\usetikzlibrary{patterns}
\usepackage{sgame}
\usepackage{color}
\usepackage{hyperref}
\usepackage{times}
\usepackage{enumitem}
\usepackage{csquotes}
\usepackage{multirow,array}
\usepackage[top=1in, bottom=1in, left=1in, right=1in]{geometry}
\begin{document}
\numberwithin{equation}{section}



\maketitle
\noindent \textbf{Abstract:} I study repeated games with anonymous random matching where players endogenously decide whether to disclose signals about their past actions. I establish an-anti folk theorem, that when players are sufficiently long-lived, they will almost
always play their dominant actions and will almost never cooperate. When players' expected lifespans are intermediate, they can sustain some cooperation if their actions are substitutes but cannot sustain any cooperation if their actions are complements. Therefore, the maximal level of cooperation a community can sustain is not monotone with respect to its members' expected lifespans and the complementarity of players' actions can undermine their abilities to sustain cooperation.\\



\newtheorem{Proposition}{\hskip\parindent\bf{Proposition}}
\newtheorem{Theorem}{\hskip\parindent\bf{Theorem}}
\newtheorem{Lemma}{\hskip\parindent\bf{Lemma}}
\newtheorem{Corollary}{\hskip\parindent\bf{Corollary}}
\newtheorem*{Definition}{\hskip\parindent\bf{Definition}}
\newtheorem*{Assumption}{\hskip\parindent\bf{Regular Records}}
\newtheorem{Condition}{\hskip\parindent\bf{Condition}}
\newtheorem{Claim}{\hskip\parindent\bf{Claim}}
\newtheorem*{Assumption1}{\hskip\parindent\bf{Assumption 1'}}

\begin{spacing}{1.5}
\section{Introduction}\label{sec1}
How can we incentivize a group of selfish individuals to take cooperative actions? Can cooperation be sustained when people interact with different partners over time? These classic questions in economics motivated the literature on \textit{community enforcement}. In communities with relatively few players, Kandori (1992), Ellison (1994), and Deb, Sugaya and Wolitzky (2020) show that players can cooperate even when they have no information about others' histories. In communities with \textit{a large number of players}, which are usually modeled as a continuum of players, sustaining cooperation requires players to have some information about their partners' histories (Takahashi 2010). Such information is called a player's \textit{record}, which may consist of signals about his past actions and possibly also signals about his previous partners' actions.

This paper contributes to the literature on community enforcement by studying situations where players can endogenously decide which signals to include in their records. One motivation comes from online platforms such as Yelp where restaurants
sometimes bribe consumers for erasing negative reviews, by which they can affect the set of reviews future consumers observe.\footnote{For example, Tadelis (2016) shows that sellers' bribes and harassment against consumers who left negative reviews have caused significant biases in online ratings. Livingston (2005) finds that negative reviews are rare on eBay and sellers' sales depend mostly on the number of good reviews. Nosko and Tadelis (2015) document that only 0.07\% of the reviews on eBay are negative despite their survey shows that a much larger fraction of the consumers are dissatisfied and complained to consumer service.}
I show that players cannot sustain any cooperation when they are \textit{sufficiently long-lived} or sufficiently short-lived, but can sustain some cooperation when their expected lifespans are \textit{intermediate}. These cooperative equilibria are robust to private payoff information when players' actions are strategic substitutes but not when their actions are strategic complements.

For an illustration of my model and results, consider an example with two populations of players. Each period, they are matched into pairs, one from each population, uniformly at random to play a two-player game (e.g., the prisoner's dilemma, the product choice game, etc.).
Each player's action generates a \textit{signal}.

My modeling innovation is that after each player observes the realized signal about his action, he can choose whether to \textit{erase} that signal or to \textit{disclose} it.\footnote{My results extend to the case where players face a cost of erasing signals, as long as that cost is small enough. This corresponds to situations in which sellers need to pay a small bribe (e.g., a 10 dollar giftcard)  to buyers in order to erase their negative reviews.} A player's \textit{record} consists of all the signals he disclosed. Each player can observe his current partner's record, but \textit{cannot} directly observe the number of times that his partner has played the game and the exact dates at which his partner's disclosed signals (if any) were generated. This information structure is motivated by some online platforms such as Yelp for restaurant reviews: Yelp shows in salient places the total number of ratings a restaurant has received, the restaurant's average rating, and the number of times that it has received each rating (1 to 5 stars). However, Yelp does not reveal how many people have visited the restaurant and it is time-consuming for most consumers to learn the exact date of each rating, especially when there is a large number of ratings.\footnote{My model can accommodate situations in which some players \textit{cannot} erase signals (e.g., buyers cannot erase negative reviews about themselves) as well as settings where sellers \textit{cannot} rate buyers, in which case the buyer's signal space is a singleton.}

By the end of each period, a constant fraction of players from each population irreversibly exit the game and are replaced by the same mass of new players. Then all the current matches are dissolved. In the next period, all the players who are still active will be matched with new partners, uniformly at random.

My first result (Theorem \ref{Theorem1}) is an anti-folk theorem, which shows that if players in a population have a strictly dominant action in the stage game and are \textit{sufficiently long-lived},\footnote{In my model, a constant fraction $1-\overline{\delta}_i$ of players from each population $i$ exit the game after each period. Therefore, the expected lifespan of each player in population $i$ is $(1-\overline{\delta}_i)^{-1}$.} then they will almost always play their dominant action in all equilibria. This result allows players to be arbitrarily patient and applies independently of the other population's payoff, discount factor, and ability to erase signals.
It is also robust to noise in players' records as well as players' misinterpretations of others' records.
In the prisoner's dilemma, my anti-folk theorem implies that sufficiently long-lived players will almost always defect. In the product choice game, my  result implies that sufficiently long-lived sellers will almost never supply good products.
In games where players do not have strictly dominant actions, a corollary of my theorem implies that sufficiently long-lived players will almost always take rationalizable actions.

Intuitively, each player has the option to sustain his current record by erasing his signal,
which implies that his \textit{continuation value} is \textit{non-decreasing} over time.
Hence, a player will take his dominant action unless taking other actions can significantly increase his continuation value.
This leads to an upper bound on the number of times that a player can be motivated to take other actions (i.e., cooperate).
It also implies that players will have strict incentives to take their dominant actions
when their continuation values reach their maximum. When players are sufficiently long-lived, \textit{either} they will cooperate with low probability when they have low continuation values, \textit{or} a large fraction of them will reach their highest continuation value after which they will never cooperate. Both cases result in a low average probability of cooperation.

Theorem \ref{Theorem1} suggests that sufficiently long-lived players will almost never cooperate. Sufficiently short-lived players
have no incentive to cooperate since their discount factors are too low. A natural question is: Can players
sustain some cooperation in \textit{some equilibria} when their expected lifespans are \textit{intermediate}?

Motivated by the practical concerns that mixed-strategy equilibria might be hard to interpret and may not be robust when
players have private information about their own payoffs,\footnote{The relevance of this robustness concern is well-explained in Bhaskar, Mailath and Morris (2013), that economic models are approximations of reality, in which case their specifications of players' payoffs may not be exactly the same as players' true payoffs. Moreover, it is conceivable that in practice, people will always have some private information about their own preferences.} I examine whether there exist cooperative equilibria that are \textit{purifiable}
as in Bhaskar (1998) and Bhaskar and Thomas (2019). The cooperative equilibria I construct will also satisfy other robustness criteria,  such as robustness to noise in the monitoring technology and robustness to players' misinterpretations of other players' records.


Theorem \ref{Theorem2} shows that regardless of the time preferences and survival probabilities, players will never cooperate in any purifiable equilibrium when their opponents have weakly supermodular payoffs,\footnote{In some supermodular prisoner's dilemma game, players can sustain some cooperation in some \textit{non-purifiable} equilibria.} i.e., their opponents find it weakly less costly to cooperate when they cooperate. In the product choice game, the buyer's payoff is supermodular since they have stronger incentives to buy larger quantities when the seller supplies good products, in which case my theorem implies that the seller will never supply good products in any purifiable equilibrium. In the prisoner's dilemma, my theorem implies that players will never cooperate in any purifiable equilibrium as long as it is weakly less costly to cooperate when others cooperate.

In contrast, Theorem \ref{Theorem3} shows that as long as both populations of players are not too impatient and have \textit{strictly submodular} stage-game payoffs, there exists an \textit{intermediate range} of survival probabilities under which players can sustain a strictly positive level of cooperation in some purifiable equilibria.

The comparison between Theorems \ref{Theorem1} and \ref{Theorem3} suggests that the maximal level of cooperation a community can sustain is \textit{not monotone} with respect to the expected lifespans of its members. This conclusion stands in contrast to most of the existing literature on repeated games where higher effective discount factors (which is the case when players have longer expected lifespans) cannot decrease the maximal level of cooperation.

The comparison between Theorems \ref{Theorem2} and \ref{Theorem3} suggests that the complementarity of players' actions can undermine their abilities to sustain cooperation. This conclusion stands in contrast to the ones obtained from models where players cannot erase signals, such as Takahashi (2010), Heller and Mohlin (2018), and Clark, Fudenberg and Wolitzky (2021), which show that when records are first order, players can sustain cooperation when their actions are strategic complements but not when their actions are strategic substitutes.

Intuitively, take the prisoner's dilemma and compare any player who is supposed to receive the \textit{highest} continuation value with anyone from the same population who is supposed to receive the \textit{lowest} continuation value. Due to the ability to erase signals,
players with the highest continuation value will always defect. When actions are complements, any player who has an incentive to cooperate with people who always defect will also have an incentive to cooperate with anyone else. Therefore, in any purifiable equilibrium and for any player, the probability that he cooperates with someone who always defect (e.g., players with the highest continuation value) \textit{cannot} be greater than the probability that he cooperates with players with the lowest continuation value.
As a result, players with the highest continuation value \textit{cannot} receive a strictly higher payoff than ones with the lowest continuation value. This rules out all the intertemporal incentives.

When actions are substitutes, players have stronger incentives to cooperate with opponents who defect. This allows the society to credibly deliver high continuation values to players who have cooperated before. My proof uses this idea and constructs a class of cooperative equilibria in which players in each population are divided into two groups: \textit{juniors} who have not cooperated before and \textit{seniors} who have cooperated at least once. Seniors defect against everyone. Juniors cooperate with seniors for sure and cooperate with other juniors with probability less than one. This is an equilibrium when players' expected lifespans are \textit{intermediate} and it is purifiable since players have strict incentives except when juniors are matched with each other.
However, this strategy profile is \textit{not} an equilibrium when players are sufficiently long-lived since there will be too many seniors in the population, which makes becoming seniors unattractive for juniors.

I review the related literature in the remainder of this section. Section \ref{sec2} sets up a general model in which players' signals can depend on
their own actions and records as well as the actions and records of their partners. It also allows for general record transition rules, various types of noise in record-keeping, as well as players misinterpreting others' records.
Section \ref{sec3} shows that
sufficiently long-lived players will almost never cooperate under a regularity condition on the record system. Section \ref{sec4}
examines the possibility of cooperation when players have intermediate expected lifespans.
Section \ref{sec5} concludes and discusses alternative modeling assumptions and record disclosure rules, such as (i) each player can disclose any subset of his past signals and (ii) each player can disclose
at most $K$ past signals.

\paragraph{Related Literature:} This paper contributes to the literature on community enforcement pioneered by Kandori (1992) and Ellison (1994), and is recently reviewed by Wolitzky (2022). It is most closely related to community enforcement models with a large number of players (usually modeled as a continuum of players), such as Takahashi (2010), Heller and Mohlin (2018), Bhaskar and Thomas (2019), and Clark, Fudenberg and Wolitzky (2021) in which observing other players' records is \textit{necessary} to sustain cooperation.\footnote{In contrast to games with a finite number of players, players cannot sustain cooperation via contagion strategies  when there is a continuum of players. This is because the probability that a player is matched again with his current partner (or being matched with any player his partner has been or will be matched with) is zero. Levine and Pesendorfer (1995) provide conditions under which the equilibrium outcomes of games with a large number of players converge to those of games with a continuum of players.}

Compared to those papers, I introduce the possibility of players strategically disclosing their signals. I show that the endogenous disclosure of signals rules out cooperation \textit{either} when players are sufficiently long-lived \textit{or} when their actions are complements. However, they can still sustain cooperation when their actions are substitutes and their expected lifespans are intermediate. These conclusions stand in contrast to the ones in the existing literature that higher effective discount factors and strategic complementarities facilitate cooperation.\footnote{Wiseman (2017) and Sandroni and Urgun (2018) also show that higher effective discount factors can undermine cooperation. In contrast to the current paper that focuses on repeated games, their results are obtained in \textit{stochastic games with absorbing states}.} My analysis also highlights the different strategic implications of players' \textit{time preference} and their \textit{survival probability}, which to the best of my knowledge, is novel in the community enforcement literature. By contrast, the two are equivalent in most of the existing repeated game models since the equilibrium outcomes depend only on the product of the two, that is, the effective discount factor.

Ghosh and Ray (1996) and Fujiwara-Greve and Okuno-Fujiwara (2009) study repeated games with voluntary separation where players can interact with the same partner in multiple periods as opposed to being rematched after each period. In their models, each player's outside option is his continuation value from joining the unmatched pool whereas in my model, his outside option is his current continuation value.

Friedman and Resnick (2001) study repeated prisoner's dilemma without any complementarity or substituability and each player can \textit{either} disclose all signals \textit{or}
erase all signals and restart with a fresh record without any signal.
In contrast, the players in my model can decide whether to disclose each signal. This gives them more flexibility in choosing what to disclose. In contrast to their results, I show that players' abilities to sustain cooperation is \textit{not monotone} with respect to their effective discount factors. Smirnov and Starkov (2022), Hauser (2023), and Sun (2024) study dynamic censoring games in which players' payoffs depend on the state. In contrast to their models, players' payoffs depend only on their actions in my model.

Pei (2023) studies a repeated game with \textit{incomplete information} where a long-lived player can erase past actions from their records. It shows a bad reputation result, which is driven by the observation that the speed of reputation building vanishes as the long-lived player's survival probability goes to one. In contrast, the current paper studies a complete information game, i.e., there is neither reputation building nor learning, but it allows for multiple long-lived players, more general stage-game payoffs, and imperfect monitoring.

The players in my model decide whether to disclose each of their  past signals but \textit{cannot} fabricate signals. This is related to the literature on disclosing hard information pioneered by Grossman (1981), Milgrom (1981), and Dye (1985). More closely related are the works of Dziuda (2011), Di Tillio, Ottaviani and S{\o}rensen (2021), and
Gao (2024) in which a sender has some pieces evidence about a payoff-relevant state and chooses a subset of them to disclose to an uninformed receiver. The unraveling results
in Grossman (1981) and Milgrom (1981)
do not apply to these models since the receiver does not know the set of evidence available to the sender. In contrast to their static models in which the distribution of the evidence available to the sender depends only on some exogenous state, my model is dynamic in which the number of each signal available to each player depends endogenously on players' behaviors.

\section{The Baseline Model}\label{sec2}
Consider a doubly infinite repeated game where time is indexed by $k=...-1,0,1,...$ There are two populations of players, indexed by $i \in I \equiv \{1,2\}$. Each period, a unit mass of players from each population are active. Unlike in standard repeated game models such as Fudenberg and Maskin (1986) where a player's \textit{time preference} and \textit{survival probability} play the same role, their roles can be different in my model since a player's survival probability affects not only his \textit{effective discount factor} but also his \textit{expected lifespan}. The latter may affect the composition of the population, which may in turn affect players' incentives.

The above concern motivates me to distinguish between the two reasons for why players discount future payoffs:
First, by the end of each period $k \in \mathbb{N}$ and before the start of period $k+1$, a fraction $1-\overline{\delta}_i$ of the active players in population $i$ \textit{irreversibly become inactive} and are replaced by the same mass of new players, with $\overline{\delta}_i \in [0,1)$. Second, conditional on remaining active in period $k+1$, each player in population $i$ is indifferent between $1$ unit of utility in period $k+1$ and $\widehat{\delta}_i \in [0,1)$ unit in period $k$.\footnote{I will comment on the case in which $\widehat{\delta}_i=1$ and the case in which $\overline{\delta}_i=1$ in Section \ref{sec5}.} Hence, each player in population $i$ has an \textit{expected lifespan} of $(1-\overline{\delta}_i)^{-1}$ and an \textit{effective discount factor} of $\delta_i \equiv \widehat{\delta}_i \cdot \overline{\delta}_i$. In standard repeated game models, the equilibrium outcomes depend on $\widehat{\delta}_i$ and $\overline{\delta}_i$ only through their product.

Each period, a fraction $p \in (0,1]$ of the active players from each population are matched into \textit{pairs} uniformly at random
to play a two-player normal form game $\mathcal{G} \equiv \{I,A,u\}$,\footnote{I allow for the case considered in the introduction that all the active players will be matched in every period, i.e., $p=1$.} 
where $A \equiv \prod_{i=1}^2 A_i$  is the set of action profiles with $A_i$ a \textit{finite} set of actions for the player from population $i$ (or \textit{player $i$}) and $u_i: A \rightarrow \mathbb{R}$ is player $i$'s stage-game payoff. Without loss of generality,
I normalize the unmatched players' and the inactive players' stage-game payoffs to $0$. Player $i$ maximizes his discounted average payoff $\sum_{k=1}^{+\infty} (1-\delta_i)\delta_i^{k-1} u_i(a_{i,k},a_{-i,k})$ where $(a_{i,k},a_{-i,k})$ is the action profile in the $k$th period of his life.

Each pair of players' actions generate a vector of signals $s \equiv (s_1,s_2)$. Player $i$'s signal is $s_i$, which is drawn from a finite set $S_i$ according to distribution $f_i (\cdot|a_i,r_i,a_{-i},r_{-i}) \in \Delta (S_i)$. This formulation allows the signal distribution $f_i$ to depend on (i) player $i$'s current-period action $a_i$ as well as that of his current matching partner's $a_{-i}$, and (ii) player $i$'s current-period \textit{record} $r_i$ as well as that of his current matching partner's $r_{-i}$.\footnote{All my results extend when each player's signal distribution depends not only on the actions and records of players in his current group but also on the joint distribution over records and actions in the rest of the population. This is because my results examine the properties of \textit{steady state equilibria}, in which the distribution over records and actions remains constant over time.}
I will formally define \textit{record} later. Each unmatched player automatically generates a \textit{null signal}, which I denote by $\emptyset$. Without loss of generality, I assume that $\emptyset \in S_i$ for every $i \in I$.

A \textit{record system} for player $i$ is $(R_i,\Psi_i^0,\Psi_i,\Psi_i^{\emptyset})$ where $R_i$ is a countable set of records with $r_i \in R_i$ a typical element, $\Psi_i^0 \in \Delta (R_i)$ is the distribution of records for new player $i$, and $\Psi_i (\cdot |r_i,s_i,r_{-i},s_{-i}) \in \Delta (R_i)$ and $\Psi_i^{\emptyset}(\cdot|r_i) \in \Delta (R_i)$  define a \textit{record transition rule} for player $i$. In particular, $\Psi_i$ is the distribution of player $i$'s record in the next period when he is matched, which can depend on the current records and signals of both players in his match, and $\Psi_i^{\emptyset}$ is the distribution of player $i$'s record in the next period when he is unmatched, which depends only on his current record (since he will generate the null signal $\emptyset$ for sure).

I introduce a regularity condition on the record system. It requires that a player's record to remain the same when he generates the null signal, which includes but not limited to the case where he is unmatched.
\begin{Definition}
Player $i$'s record system is regular if $\Psi_i (r_i|r_i,s_i=\emptyset,r_{-i},s_{-i})=1$ and
$\Psi_i^{\emptyset} (r_i|r_i)=1$
\end{Definition}
For an example of a record system that satisfies my regularity condition, consider restaurant reviews on Yelp. A restaurant generates a null signal if no consumer dined there in a certain period (i.e., it is unmatched), or when some consumers dined there (i.e., it is matched) but did not post reviews. A restaurant's record system is regular if its record consists of
(i) the number of ratings it has received,
(ii) its average rating, and (iii) the summary statistics of its ratings (e.g., the number of times that it has received each rating).
This is because all three variables remain the same when no rating is posted. I will discuss this example in more detail in Section \ref{sub2.1} and I will discuss the implications of my regularity condition in Section \ref{sub2.2}.

As in Takahashi (2010) and Clark, Fudenberg and Wolitzky (2021), after player $i$ is matched, he can observe his current partner's record $r_{-i}$ in addition to his current record $r_i$ before choosing his action $a_i$. Players \textit{cannot directly observe} any additional information about their matched partners, such as their partners' age in the game and the number of times that they were matched before. Nevertheless, in equilibrium, players can infer the values of these variables using Bayes rule after observing their partners' records.

My modeling innovation is that after player $i$ observes his realized signal $s_i$ but before his record in the next period is realized, he has the option to \textit{erase} $s_i$ by replacing it with the null signal $\emptyset$. Hence, if player $i$ is matched and does not erase $s_i$, then his record in the next period will be distributed according to $\Psi_i (\cdot |r_i,s_i,r_{-i},s_{-i})$.
When player $i$'s record system is regular,
if he is matched and erases $s_i$ or if he is unmatched, then his signal will be $\emptyset$ and his record in the next period will be the same as his record in the current period. Then a fraction $1-\overline{\delta}_i$ of player $i$ become inactive and are replaced by new players.

Player $i$'s \textit{strategy} is denoted by $\sigma_i \equiv (\sigma_i^a,\sigma_i^e)$, where $\sigma_i^a: \prod_{i=1}^2 R_i \rightarrow \Delta (A_i)$ is a mapping from his current-period record $r_i$ and the record of his current partner $r_{-i}$ to a distribution over his actions and $\sigma_i^e: R_i \times S_i  \rightarrow [0,1]$ is a mapping from his current-period record $r_i$ and his current-period signal $s_i$ to the probability with which he erases $s_i$.\footnote{Whether player $i$ observes $s_{-i}$ before deciding whether to erase $s_i$ will not affect the equilibrium outcomes. Whether player $i$ observes $s_i$ before deciding whether to erase it may affect the equilibrium outcomes but will not affect the validity of my results.} Although players may also observe their past records, the signals they erased, their previous partners' records, and so on, focusing on equilibria where no player conditions their behavior on
any payoff irrelevant information is without loss of generality in terms of players' equilibrium payoffs and the expected probabilities with which they take each of their actions.\footnote{Formally speaking, fix the parameters of the game and fix any equilibrium
under these parameters
in which players' strategies depend on any additional information, there exists an equivalent equilibrium in which (i) all players use strategies that belong to my class and (ii) their discounted average payoffs and the average probability with which they take each of their actions remain unchanged. This is because there is a continuum of players and matching is uniform, which implies that the probability with which a player will encounter one of his previous partners (or anyone that his previous partners have interacted with in the past or will interact with in the future) is zero. Hence, the equilibrium outcomes depend on a player's strategy only through (i) his expected action conditional on each $(r_i,r_{-i})$ and (ii) his expected probability of erasing each signal $s_i$ conditional on $(r_i,s_i)$.}

The solution concept is steady state equilibrium, or \textit{equilibrium} for short, which consists of a distribution over records $\mu_i \in \Delta (R_i)$ and a strategy $\sigma_i$ for each population $i \in \{1,2\}$ such that
(i) for every $i \in \{1,2\}$, $\sigma_i$ maximizes player $i$'s discounted average payoff when the record distribution is $(\mu_1,\mu_2)$ and players in the other population use strategy $\sigma_{-i}$ and (ii)  $(\mu_1,\mu_2)$ is a steady state record distribution when players behave according to $\{\sigma_i\}_{i=1}^2$. I will introduce stronger solution concepts when presenting positive results.

There exists at least one steady state equilibrium in the repeated game since (i)
there exists at least one Nash equilibrium in the stage game given that both players' action sets are finite and (ii)
all players playing a particular stage-game Nash equilibrium at all information sets is an equilibrium of the repeated game.

\subsection{Examples of Regular Record Systems}\label{sub2.1}
My regularity condition on players' record systems seems to fit some of the online review platforms such as restaurant reviews on Yelp, apartment reviews on apartmentratings.com, and so on.

For example, when a consumer uses Yelp to search for restaurants that belong to a certain category (e.g., French restaurants in NYC), Yelp
shows a \textit{list} of restaurants that belong to this category and
discloses for each restaurant on that list (i) the number of ratings it has received and (ii) its average rating (a rational number from $1$ to $5$). After a consumer clicks on a restaurant, it also shows in a salient place the number of each rating (from $1$ star to $5$ stars) the restaurant has received, that is, the \textit{summary statistics} of the ratings.

To fit this into my formulation, suppose each restaurant's signal $s_i$ is drawn from set $S_i \equiv \{\emptyset,1,2,3,4,5\}$, where $\emptyset$ stands for no review being posted and $s_i=m \in \{1,2,3,4,5\}$ stands for the restaurant receiving an $m$-star review. The distribution of $s_i$ may depend on the restaurant's action (e.g., its effort), the consumer's action (e.g., the dishes they ordered), and potentially, the restaurant's current record.
For a restaurant that has operated for $K$ periods, its record consists of
the number of each \textit{non-null} review it has received $r_{i,K} \equiv (r_{i,K}^m)_{m=1}^5$ where $r_{i,K}^m \equiv \# \{k \leq K |s_{i,k} =m\}$ for every $m \in \{1,2,3,4,5\}$. Importantly, the number of null signals is \textit{not} included in the records.
If restaurant $i$ is new, then its record is $(r_{i,0}^1,r_{i,0}^2,...,r_{i,0}^5)=(0,0,...,0)$.
The record transition rule $\Psi_i$ is deterministic with (i) $r_{i,K+1}=r_{i,K}$ when $s_{i,K}=\emptyset$, i.e., a restaurant's record remains unchanged when it is unmatched, when the consumer does not post any review, or when he is persuaded to erase his review, and (ii) when the restaurant received an $m$-star review
$s_{i,K}=m$, $r_{i,K+1}=(r_{i,K+1}^m)_{m=1}^5$ where $r_{i,K+1}^{\tau}=r_{i,K}^{\tau}$ for every $\tau \neq m$ and $r_{i,K+1}^{m}=r_{i,K}^m+1$.

A restaurant's record system also satisfies my  condition when its record consists of the number of ratings and its average rating but not the summary statistics of ratings. To see this, let us redefine the record space $R_i$ to be $\mathbb{N} \times \mathbb{Q}[1,5]$ where $\mathbb{Q}[1,5]$ is the set of rational numbers between $1$ and $5$ and then rewrite the record transition rule induced by this new record  system. A record system is also regular when the record consists of the \textit{exact sequence} of its disclosed ratings, i.e., for each pair of unerased ratings, the consumers can observe which one arrived first.
On Yelp, consumers can easily obtain such information by clicking ``Newest First'' and the system will show the number of stars in each rating starting from the latest.

\subsection{Discussions of Modeling Assumptions}\label{sub2.2}
My model distinguishes between players' time preference and survival probability. The motivation is that unlike in canonical repeated game models where these two variables play the same role and affect the equilibrium outcomes only through their product (i.e., the effective discount factor), their roles are different when players' records are affected by their disclosure decisions.\footnote{The distinction between time preference and survival probability is also noted in the literature on steady state learning, such as Fudenberg and Levine (1993), Fudenberg and He (2018), and Clark and Fudenberg (2021). Their results require players' survival probabilities to be much closer to $1$ relative to their time preferences since they need a large fraction of players to learn the steady state action distribution. This is different from the rationale behind my results that old players have no incentive to cooperate.} These two variables are also conceptually different. For example, a restaurant's survival probability captures the chances that they shut down for exogenous reasons whereas its time preference is determined by the interest rate and its owner's patience. My model is also equivalent to one where players do not exit  but instead their records are automatically erased with probability
$1-\overline{\delta}_i$ after each period. This is because players are matched uniformly and there is a continuum of players, which implies that a player's payoff and incentive depend only on their own record.

When a player's record system is regular, his record \textit{cannot} directly reveal the number of times that he has played the game and the exact date at which each of his disclosed signal was generated.\footnote{Players cannot directly observe their opponents' age in the game is a standard assumption in repeated games with limited information flows, such as those with limited memories or with voluntary separation (Fujiwara-Greve and Okuno-Fujiwara 2009).}
This is because once a player's record reveals any of the above information, his record will \textit{not} be the same after generating a null signal. My regularity condition fits online platforms such as Yelp: Yelp does not show the number times each restaurant was visited,
which makes it hard for consumers to obtain such information. Although information about the exact date of each rating is shown on Yelp, it is not displayed in salient places: Obtaining such information requires consumers to click on the restaurant, to scroll down, and to read each time stamp, which seems rather time-consuming when there are hundreds or thousands of ratings. Hence, it seems reasonable that most consumers make their decisions based on the number of ratings, the average rating, and the summary statistics of the ratings, instead of on the exact date of each rating.

My model assumes that players can erase signals without paying any cost. My results extend when players face a strictly positive cost to erase signals (e.g., restaurants need to bribe consumers to erase negative reviews) as long as that cost is small enough.
My model focuses on a particular form of record manipulation, that players can erase their signals, i.e., replacing the realized signal with the null signal. Theorems \ref{Theorem1}, \ref{Theorem2}, and \ref{Theorem4} apply as long as this form of record manipulation is feasible, even when players can also manipulate records in alternative ways, such as changing their realized signal to other non-null signals, i.e., fabricating signals. Theorem \ref{Theorem3} applies when players can also fabricate signals, as long as their cost of doing so is above some cutoff. In the prisoner's dilemma, it requires the cost of fabrication to be greater than the cost of cooperation. The assumption that the cost of fabrication to be much greater than the cost of erasing signals seems reasonable
given the observation in Tadelis (2016) that most of the consumers post reviews because they are intrinsically motivated to share their opinions, to reward sellers' good behaviors and to punish bad ones, or to provide future consumers useful information. If this is the case, then it seems much more costly for sellers to convince consumers to explicitly lie about their experiences than to ask them to stay silent.

My results extend to several alternative models of record disclosure, for example, when players can erase their signals not only in the period they generated them but also in all future periods.  They also apply to models where each player \textit{cannot} erase signals but can disclose to his partners \textit{any subset} of the signals he has generated.
They are also true when players face constraints when choosing which signals to disclose, such as they can disclose at most $K \in \mathbb{N}$ signals, as in the works of Di Tillio, Ottaviani and S{\o}rensen (2021) and Farina, Fr\'{e}chette, Lizzeri and Perego (2024). I will provide more details in Section \ref{sec5}.

Although this is not captured in the baseline model in order to avoid cumbersome notation, my results also extend to cases where players may \textit{misinterpret} others' records. This can be modeled as players observing \textit{noisy signals} about their partners' records instead of observing them perfectly. For example, when player $i$'s record is $r_i$, his partners interpret it as $\widetilde{r}_i$ where  $\widetilde{r}_i=r_i$ with probability $1-\varepsilon$ and is drawn according to $\Gamma_i (\cdot|r_i) \in \Delta (R_i)$ otherwise. Players take actions and make disclosure decisions based on
their own records, their interpretations of their current partners' records, and their realized signals.

My analysis focuses on the steady state equilibria of a doubly infinite repeated game as in Heller and Mohlin (2018) and Clark, Fudenberg and Wolitzky (2021)
as opposed to the Perfect Bayesian equilibria in repeated games where calendar time starts from $0$
(e.g., Takahashi 2010). This is because in steady states, it is straightforward to define the probabilities with which each player takes each of their actions. It also makes the analysis more tractable in the sense that players do not need to learn about the distribution over records in the population from his current partner's record, given that it is constant in the steady state.

\section{Impossibility of Cooperation Between Sufficiently Long-Lived Players}\label{sec3}
Even though players can selectively disclose signals, they may still have incentives to cooperate (i.e., to take actions that are not optimal in the stage game) when their \textit{effective discount factors} are large enough. This is because although
they can replace bad signals with the null signal, they \textit{cannot} fabricate good signals (i.e., changing a signal to some non-null signal). As a result, a large number of good signals can be generated only when a player has cooperated many times before. This seems to suggest that a player will have an incentive to take cooperative actions when he will be rewarded for disclosing many good signals.

I show that the above intuition breaks down when players are \textit{sufficiently long-lived}.
For any population $i$, given their strategy $\sigma_i$ and the steady state record distribution $(\mu_1,\mu_2)$, the \textit{average probability} with which player $i$ takes action $a_i \in A_i$ is the probability that random variable $\sum_{(r_1,r_2) \in R_1 \times R_2} \mu(r_1,r_2) \sigma_i^a (r_1,r_2)$ assigns to action $a_i$.
I establish an anti-folk theorem when players are sufficiently long-lived:
\begin{Theorem}\label{Theorem1}
Suppose there exists $i \in I$ such that player $i$ has a regular record system as well as a strictly dominant action $a_i^* \in A_i$ in the stage game. Then for every $\widehat{\delta}_i \in (0,1)$ and $\varepsilon>0$, there exists $\delta^* \in (0,1)$ that depends only on
$(\widehat{\delta}_i,u_i)$ such that when $\overline{\delta}_i > \delta^*$, the average probability with which player $i$ takes action $a_i^*$ is greater than $1-\varepsilon$ in every equilibrium.
\end{Theorem}
Theorem \ref{Theorem1} implies that as long as players in a population are sufficiently long-lived and have regular record systems, they will almost never cooperate and will almost always take their strictly dominant action in all equilibria. When the stage game is the prisoner's dilemma, my theorem implies that sufficiently long-lived players will almost always defect in all equilibria.  When buyers and sellers play the product choice game, my theorem implies that sufficiently long-lived sellers will almost never supply good products.

My anti-folk theorem applies even when (i) player $i$ is patient (i.e., $\widehat{\delta}_i$ can be close to $1$) and (ii) player $i$'s effective discount factor $\delta_i$ is arbitrarily close to $1$, which is the case when both $\widehat{\delta}_i$ and $\overline{\delta}_i$ go to $1$ but $\overline{\delta}_i$ goes to $1$ at a faster rate relative to $\widehat{\delta}_i$. This theorem also holds independently of other players' record systems (e.g., whether their record systems are regular), stage-game payoff functions, survival probabilities, and time preferences, as well as whether other players have the ability to erase their signals. It also extends to environments where there are $n \geq 2$ populations of players who are being randomly matched into groups of size $n$ in each period to play an $n$-player finite normal form game.

The takeaway from Theorem \ref{Theorem1} is that sufficiently long-lived players cannot sustain cooperation when they selectively disclose past signals. This stands in contrast to the standard intuition  from canonical repeated game models in which fixing a player's time preference, an increase in his expected lifespan leads to a higher effective discount factor, which can strengthen this player's incentive to sacrifice his current-period payoff in exchange for a higher continuation value in the future and can lead to more cooperation.

The intuition behind my result is that when a player has the option to erase signals from his records, his \textit{expected} continuation value must be \textit{non-decreasing} over time. As a result, in order to motivate player $i$ to cooperate, that is, to take any action $a_i$ that is not his strictly dominant action $a_i^*$,
his expected continuation value needs to increase by at least something proportional to $1-\delta_i$. This implies that
\textit{old players} whose continuation values are close to their highest level will have strict incentives to play $a_i^*$. When players are sufficiently long-lived, \textit{either} they cooperate with low probabilities when their continuation values are low, \textit{or} they cooperate with probabilities that are bounded above zero when they are young, in which case there will be a large fraction of them reaching their highest continuation value after which they will have no incentive to cooperate. In both cases,
the average probability with which player $i$ cooperates is low.

Although an increase in player $i$'s survival probability $\overline{\delta}_i$ can also increase his effective discount factor $\delta_i$, which can strengthen his incentive to cooperate, this effect is  dominated since the expected number of periods that player $i$ may have an incentive to cooperate is at most proportional to $(1-\delta_i)^{-1}$. These periods are negligible relative to player $i$'s expected lifespan
$(1-\overline{\delta}_i)^{-1}$ when the ratio $\frac{1-\overline{\delta}_i}{1-\delta_i}$ vanishes to $0$.

\begin{proof}[Proof of Theorem 1:] Fix any steady state equilibrium $(\sigma,\mu)$.
Let $V(r_i)$ denote player $i$'s expected continuation value when his record is $r_i$ \textit{before} knowing his current match.
Let $R_i^* \subset R_i$ denote the set of player $i$'s records that occur with strictly positive probability under $\mu$.
Let $\overline{V} \equiv \sup_{r_i \in R_i^*} V(r_i)$ and
$\underline{V} \equiv \inf_{r_i \in R_i^*} V(r_i)$
denote player $i$'s highest and lowest continuation values, respectively.
Let $\alpha_{-i}(r_i,r_{-i}) \in \Delta (A_{-i})$ denote player $i$'s expectation of his opponent's action when his record is $r_i$ and his opponent's record is $r_{-i}$.
Let $f_i(s_i|a_i,a_{-i},r_i,r_{-i})$ denote player $i$'s expected probability of generating signal $s_i$ conditional on the current-period action profile $(a_i,a_{-i})$ and record profile $(r_i,r_{-i})$. When player $i$ with record $r_i$ is matched with a partner with record $r_{-i}$, he has an incentive to take action $a_i' \neq a_i^*$ only if
\begin{equation*}
(1-\delta_i) u_i(a_i',\alpha_{-i}(r_i,r_{-i}))  + \delta_i \sum_{s_i \in S_i} f_i(s_i|a_i',r_i,\alpha_{-i}(r_i,r_{-i}), r_{-i})
\max \Big\{
V(r_i),
\sum_{r_i' \in R_i} \Psi_i (r_i'|r_i,s_i,r_{-i},s_{-i}) V(r_i')
\Big\}
\end{equation*}
\begin{equation}\label{3.1}
\geq (1-\delta_i) u_i(a_i^*,\alpha_{-i}(r_i,r_{-i}))  + \delta_i V(r_i),
\end{equation}
where the RHS is player $i$'s discounted average payoff when he plays $a_i^*$ and erases every signal that he generates.
Let
\begin{equation}\label{3.2}
c^* \equiv \min_{a_i \neq a_i^*, a_{-i} \in A_{-i}} \Big\{
u_i (a_i^*,a_{-i})- u_i (a_i,a_{-i})
\Big\} >0,
\end{equation}
which is player $i$'s lowest stage-game cost for not taking his strictly dominant action $a_i^*$.
Let $V_k$ denote player $i$'s \textit{expected} continuation value in the $k$th period of his life conditional on him being active for at least $k$ periods.
Let $\pi_k$ denote player $i$'s \textit{expected} probability of playing actions other than $a_i^*$ in the $k$th period of his life (conditional on him being active for at least $k$ periods).
Since player $i$'s expected continuation value in the next period is $\max \Big\{
V(r_i),
\sum_{r_i' \in R_i} \Psi_i (r_i'|r_i,s_i,r_{-i},s_{-i}) V(r_i')
\Big\}$ when his current-period record is $r_i$ and he generates signal $s_i$, his incentive constraint in (\ref{3.1}) leads to a lower bound on $V_{k+1}-V_k$:
\begin{equation}\label{3.3}
V_{k+1}-V_k \geq \frac{1-\delta_i}{\delta_i} c^* \pi_k.
\end{equation}
Since $\underline{V} \leq V_k \leq \overline{V}$ and $\overline{V},\underline{V}$ are bounded above and below by player $i$'s highest and lowest stage-game payoffs, denoted by $\overline{u}$ and $\underline{u}$, respectively, one can obtain from (\ref{3.3}) that
\begin{equation}\label{3.4}
\frac{1-\delta_i}{\delta_i} c^* \sum_{k=0}^{+\infty} \pi_k \leq \sum_{k=0}^{+\infty} ( V_{k+1}-V_k)
\leq
\overline{V}-\underline{V} \leq \overline{u}-\underline{u}.
\end{equation}
Since player $i$ becomes inactive with probability $1-\overline{\delta}_i$ after each period, the fraction of player $i$ with age $k$ equals $(1-\overline{\delta}_i) \overline{\delta}_i^k$. By the law of total probabilities, the average probability with which player $i$ takes actions other than $a_i^*$ is $\sum_{k=0}^{+\infty} (1-\overline{\delta}_i) \overline{\delta}_i^k \pi_k$. Applying summation by parts and using (\ref{3.4}), we obtain that
\begin{equation}\label{3.5}
\sum_{k=0}^{+\infty} (1-\overline{\delta}_i) \overline{\delta}_i^k \pi_k= (1-\overline{\delta}_i)^2 \sum_{k=0}^{+\infty} \overline{\delta}_i^k \sum_{t=0}^k \pi_t \leq (1-\overline{\delta}_i)^2 \sum_{k=0}^{+\infty} \overline{\delta}_i^k \frac{\overline{u}-\underline{u}}{c^*} \cdot \frac{\delta_i}{1-\delta_i}=\frac{1-\overline{\delta}_i}{1-\delta_i} \frac{(\overline{u}-\underline{u}) \delta_i}{c^*}.
\end{equation}
By definition, $\delta_i \equiv \overline{\delta}_i \widehat{\delta}_i$. Fix any $\widehat{\delta}_i \in (0,1)$ and let $\overline{\delta}_i \rightarrow 1$, the RHS of (\ref{3.5}) will vanish to $0$.
\end{proof}

Theorem \ref{Theorem1} applies to games where at least one player has a strictly dominant action in the stage game. Corollary \ref{Cor1} extends its findings to games without strictly dominant actions. It shows that when all players have regular record systems and have sufficiently long expected lifespans, 
they will almost always take actions that can survive iterative deletion of strictly dominated strategies in all equilibria. This finding is useful for games where players do not have strictly dominant actions but some of their actions are strictly dominated,
such as Cournot and Bertrand games where players choose prices and quantities from finite sets.
\begin{Corollary}\label{Cor1}
Suppose both players have regular record systems. For every $\widehat{\delta}_1,\widehat{\delta}_2 \in (0,1)$ and $\varepsilon>0$, there exists $\delta^* \in (0,1)$ that depends only on
$(\widehat{\delta}_1,\widehat{\delta}_2,u_1,u_2)$ such that when $\overline{\delta}_1,\overline{\delta}_2 > \delta^*$, the average probability with which both players take rationalizable actions is greater than $1-\varepsilon$ in every equilibrium.
\end{Corollary}
The proof follows from that of Theorem \ref{Theorem1} and is relegated to Appendix \ref{secA}.


\section{Cooperation Between Players with Intermediate Expected Lifespans}\label{sec4}
Theorem \ref{Theorem1} implies that players with sufficiently long expected lifespans will almost never cooperate. Players with short expected lifespans have no incentive to cooperate since their effective discount factors are too low. This leaves open the question that whether players can sustain cooperation when their expected lifespans are \textit{intermediate}, which is what I am going to examine in the current section.

I start from introducing some regularity conditions on players' payoffs since they will be used in some of my results. These conditions are standard in the community enforcement literature, which include the prisoner's dilemma game in Takahashi (2010), Heller and Mohlin (2018), and Clark, Fudenberg and Wolitzky (2021) and the product choice game in Bhaskar and Thomas (2019).
I start from a monotonicity condition:
\begin{Definition}
Player $i$'s payoff is monotone
under complete orders $\succsim_1$ and
$\succsim_2$
on $A_1$ and $A_2$ if
$u_i(a_i,a_{-i})$ is strictly increasing in $a_{-i}$ and is strictly decreasing in $a_i$.
\end{Definition}
Intuitively, a player's payoff is monotone under some complete orders on players' action sets if his ordinal preference over his actions is independent of which action his opponent takes and his ordinal preference over his opponent's action is independent of which action he takes. For example, both players' payoffs are monotone in the prisoner's dilemma game under the order $C \succ D$:
 \begin{center}
\begin{tabular}{| c | c | c |}
\hline
 - & \textbf{C}ooperate & \textbf{D}efect  \\
  \hline
  \textbf{C}ooperate & $1,1$ & $-l_1,1+g_2$ \\
  \hline
  \textbf{D}efect & $1+g_1,-l_2$ & $0,0$  \\
  \hline
\end{tabular} $\quad$ with  $g_1,g_2,l_1,l_2 >0$.
\end{center}
In the product choice game, the seller's payoff is monotone under orders $G \succ_1 B$ and $L \succ_2 S$ but the buyer's payoff is not monotone under any complete orders on players' action sets:
\begin{center}
\begin{tabular}{| c | c | c |}
  \hline
 seller $\backslash$ buyer & \textbf{L}arge Quantity & \textbf{S}mall Quantity \\
  \hline
 \textbf{G}ood Products & $1, 1$ & $-l, x$ \\
  \hline
  \textbf{B}ad Products & $1+g, -x$ & $0,0$ \\
  \hline
\end{tabular} $\quad$ with $g,l>0$ and $x \in (0,1)$.
\end{center}

If player $i$'s payoff is monotone under complete order $\succsim_i$, then (i) his lowest action  is his strictly dominant action in the stage game, which I denote by $a_i^*$, and (ii) higher actions from his opponent (i.e., player $j$) are more \textit{cooperative} in the sense that they will increase player $i$'s stage-game payoff. However, when player $j$'s payoff is also monotone, choosing a higher $a_j$ will lower his own stage-game payoff.

Next, I introduce the definitions of weakly supermodular payoffs and strictly submodular payoffs, which are standard in the literature on community enforcement and monotone comparative statics:
\begin{Definition}
Under complete orders $\succsim_1$ and $\succsim_2$ on $A_1$ and $A_2$:
\begin{itemize}
\item[1.] Player $i$'s payoff is weakly supermodular if
$u_i(a_i,a_{-i})$ has weakly increasing differences, that is,
$u_i(a_i,a_{-i})- u_i(a_i,a_{-i}')
\geq u_i(a_i',a_{-i})- u_i(a_i',a_{-i}')$ for every  $a_i \succ_i a_i'$  and $a_{-i} \succ_{-i} a_{-i}'$.
\item[2.] Player $i$'s payoff is strictly submodular if $u_i(a_i,a_{-i})$ has strictly decreasing differences, that is,
$u_i(a_i,a_{-i})- u_i(a_i,a_{-i}')
< u_i(a_i',a_{-i})- u_i(a_i',a_{-i}')$ for every  $a_i \succ_i a_i'$  and $a_{-i} \succ_{-i} a_{-i}'$.
\end{itemize}
\end{Definition}
Intuitively, a player's payoff is supermodular when he has stronger incentives to take higher actions (i.e., more cooperative actions) when his opponent's action is higher. In another word, his cost of cooperation is lower when his opponent also cooperates. A player's payoff is submodular when he has stronger incentives to take more cooperative actions when his opponent takes less cooperative actions.

In the prisoner's dilemma, under the order $C \succ D$ on players' action sets, player $i$'s payoff is weakly supermodular when
$g_i \leq l_i$ and is strictly submodular when $g_i >l_i$. My general formulation allows one player's payoff to be supermodular while the other player's payoff to be submodular. In the product choice game, under the orders $G \succ_1 B$ and $L \succ_2 S$ on players' action sets,
the buyer's payoff is always weakly supermodular since she has stronger incentives to buy larger quantities when the seller supplies good products with higher probability. The seller's payoff is weakly supermodular when $g \leq l$ (i.e., supplying good products is less costly when the buyer buys a large quantity) and is strictly submodular when $g>l$.

My results in this section examine whether players can sustain cooperation in \textit{purifiable equilibrium}, which is a refinement of
steady state equilibrium. This refinement is motivated by the concerns that mixed-strategy equilibria might be hard to interpret and they may not be robust when players have private payoff information, both of which seem relevant in practice.\footnote{In practice, there are other concerns regarding the robustness of equilibria such as the robustness to noise in the monitoring technology, the robustness to noise in players' observations, and the robustness to misinterpretations of other players' records. The cooperative equilibria I constructed in the proof of Theorem \ref{Theorem3} are robust to these concerns.} The concept of purifiable equilibria was first introduced by Harsanyi (1973) and was incorporated into the analysis of dynamic games by the works of Bhaskar (1998), Bhaskar, Mailath and Morris (2013), and Bhaskar and Thomas (2019). Unlike in static games where all equilibria are purifiable under generic payoffs, there are many mixed-strategy equilibria in dynamic games that are \textit{not} purifiable, as demonstrated by a well-known example in Bhaskar (1998).

My definition of purifiable equilibria in repeated games follows from that in Bhaskar and Thomas (2019). Recall that my stage game is a finite normal-form game $\Gamma \equiv \{I,A,u\}$, which I refer to as the \textit{unperturbed stage game}. For every $\varepsilon>0$,
an $\varepsilon$-\textit{perturbed stage game} is denoted by $\Gamma(\varepsilon)$, in which the set of players and the set of actions are the same as in the unperturbed stage game but player $i$'s payoff from playing $a_i$ is given by $u_i(a_i,a_{-i})+ \varepsilon z_i (a_i)$, where $z_i(a_i)$ is a random variable that is interpreted as a payoff shock that affects player $i$'s payoff from playing $a_i$. As in Harsanyi (1973), the distribution of $z_i(a_i)$ has bounded support and no atom, and that these random variables are independently distributed across actions, players, and periods. Each period, before player $i$ acts, he observes the realizations of his own payoff shocks $\{z_i(a_i)\}_{a_i \in A_i}$ for the current period but \textit{not} the realized payoff shocks of the other players and those in the future. An equilibrium $(\sigma,\mu)$ of the unperturbed repeated game is \textit{purifiable} if for every sequence $\varepsilon \rightarrow 0$, there exist a sequence of equilibria $(\sigma(\varepsilon),\mu(\varepsilon))$ of the  repeated games with $\varepsilon$-perturbed stage games
$\Gamma(\varepsilon)$ that converge to $(\sigma,\mu)$.

I start from a result which shows that the complementarity in players' actions is \textit{not} conducive to cooperation in the sense that players will \textit{always} take their strictly dominant actions in \textit{all} purifiable equilibria. It requires one player's record to be \textit{first order}, that is, his signal distribution and record transition rule depend only on his own actions, signals, and records, but not on other players' actions, signals, and records.
\begin{Definition}
Player $i$'s record is first-order if the distribution of his signal $f_i(\cdot|a_i,s_i,a_{-i},s_{-i})$ depends only on $(a_i,s_i)$
and his record transition rule $\Psi_i (\cdot|r_i,s_i,r_{-i},s_{-i})$ depends only on $(r_i,s_i)$.
\end{Definition}
First-order record is one of the primary focuses of the community enforcement literature. For example, they were the focus of the main results in Takahashi (2010), Heller and Mohlin (2018), and Clark, Fudenberg and Wolitzky (2021). As an example, a restaurant's record is first order as long as its customers' ratings depend only on the restaurant's action (i.e., its effort) and its current rating, but not on its customers' actions and records. On platforms where sellers cannot rate buyers, buyers' records are automatically first order since they are constant over time. On platforms where sellers can rate buyers, buyers' records are also first order as long as the ratings they receive depend only on their own behaviors and their own past records.
\begin{Theorem}\label{Theorem2}
Suppose player $1$'s record system is regular, player $2$'s record system is first order, and there exist complete orders $\succsim_1$ and $\succsim_2$ on $A_1$ and $A_2$ under which player $1$'s payoff is monotone and player $2$'s payoff is weakly supermodular, then
in all purifiable equilibria,
player $1$ will always play his strictly dominant action $a_1^*$ and player $2$ will always play some stage-game best reply to $a_1^*$.
\end{Theorem}
I will prove this result in Appendix \ref{secB} and will discuss its intuition after stating Theorem \ref{Theorem3}.
Theorem \ref{Theorem2} suggests that regardless of players' time preferences and survival probabilities,
it is impossible to motivate a player (e.g., player $1$) to take any action other than his strictly dominant action $a_1^*$
in any purifiable equilibrium  as long as (i) this player has monotone payoffs and can sustain his current record by erasing his signals, (ii) his opponent's (i.e., player $2$) record does not depend on player $1$'s actions and records, and (iii) his opponent finds it weakly more costly to reward player $1$ when player $1$ takes action $a_1^*$. I explain the implications of Theorem \ref{Theorem2} using the two leading examples. 
In the product choice game, 
\begin{enumerate}
  \item the sellers' payoffs are \textit{monotone} since they find it costly to supply good products,
  \item the buyers' payoffs are \textit{weakly supermodular} since they have stronger incentives to trust sellers who supply good products.
\end{enumerate}
Therefore, Theorem \ref{Theorem2} applies as long as 
\begin{enumerate}
\item sellers can sustain their current records by erasing their signals (i.e., their records are regular),
\item buyers' records are first order, which is the case when sellers cannot rate them or when  
sellers can rate them but the ratings the buyers receive depend only on their own actions and records.
\end{enumerate}
Theorem \ref{Theorem2} does \textit{not} require the seller's record to be first order, that is, the buyer's reviews on the seller \textit{can} depend on the buyers' actions and records. It also does \textit{not} rely on buyers' abilities to erase their signals and does \textit{not} require buyers' record systems to be regular.\footnote{Pei (2023) shows that in a repeated complete information game where the buyers are myopic, the seller will never supply good products in any equilibrium of the repeated product choice game. In contrast, Theorem \ref{Theorem2} allows the buyers to have arbitrary time preferences, survival probabilities, and allows for a larger class of monitoring technologies and record transition rules.}

Theorem \ref{Theorem2} also applies to the prisoner's dilemma game when (i) one of the players (e.g., player $2$) has first-order records and finds it weakly less costly to cooperate when his opponent cooperates and (ii) his opponent (i.e., player $1$) can sustain his current record by erasing his signals. Under these conditions, my theorem implies that cooperation will break down in the sense that in all purifiable equilibria, player $1$ will always defect, and as a result, player $2$ will always defect as well.

Focusing on games where both player roles have monotone payoffs as well as regular and first-order record systems, such as the prisoner's dilemma where players can erase signals and their signal depend only on their own action and record, Theorem \ref{Theorem2} suggests that as long as \textit{one player} finds it weakly less costly to cooperate when their opponent cooperates, there will be no cooperation in any purifiable equilibrium regardless of the time preferences and survival probabilities. A natural follow-up question is whether players can sustain cooperation in some equilibria when \textit{both} players have strictly submodular payoffs, i.e., both find it less costly to cooperate when their opponents defect. Theorem \ref{Theorem3} shows that the answer is yes as long as players are not too impatient, have \textit{intermediate} expected lifespans, and can benefit from cooperation.

For Theorem \ref{Theorem3}, consider a setting in which for every $i \in \{1,2\}$, player $i$'s signal
$s_i$ reveals $a_i$, player $i$'s record consists of the sequence of his unerased non-null signals,\footnote{This result also extends when player $i$'s record only reveals the number of times that he has generated each non-null signal.}, i.e., player $i$'s record is allowed to contain more information than that and is not required to be first order,
and there exist complete orders $\succsim_1$ and $\succsim_2$ on $A_1$ and $A_2$ under which both players' stage-game payoffs are \textit{monotone} and \textit{strictly submodular}.
I also assume that \textit{cooperation is profitable} in the sense that there exists an action profile $(\overline{a}_1,\overline{a}_2) \neq (a_1^*,a_2^*)$ such that $u_1(\overline{a}_1,\overline{a}_2)>u_1(a_1^*,a_2^*)$
and $u_2(\overline{a}_1,\overline{a}_2)>u_2(a_1^*,a_2^*)$. This requirement is satisfied in the prisoner's dilemma since mutual cooperation Pareto dominates mutual defect.
It is
not redundant for sustaining cooperation since without this condition, players will have no incentive to cooperate for a trivial reason, that their payoffs from cooperation are no greater than their  payoffs in the unique stage-game Nash equilibrium.
\begin{Theorem}\label{Theorem3}
In settings described by the above paragraph, there exists $\delta^* \in (0,1)$ such that when $\widehat{\delta}_1,\widehat{\delta}_2> \delta^*$, there exists a non-empty interval $[\delta',\delta''] \subset (0,1)$ such that when $\overline{\delta}_1,\overline{\delta}_2 \in [\delta',\delta'']$, there exists a purifiable equilibrium in which both $\overline{a}_1$ and $\overline{a}_2$ are played with strictly positive probability.
\end{Theorem}
The proof is in Appendix \ref{secC}. Theorem \ref{Theorem3} implies that the substitutability in players' actions is conducive to cooperation. In the prisoner's dilemma where both players have strictly submodular payoffs (i.e., $g_i > l_i$ for every $i \in \{1,2\}$), my result implies that players can sustain some cooperation when their expected lifespans are intermediate. A similar conclusion applies to some games where players' stage-game payoffs violate monotonicity but still satisfy strict submodularity. For example, one can use a similar construction to show that players can sustain some cooperation in Cournot games. The details are available upon request.

Although Theorem \ref{Theorem3} is stated in the case where (i) players' signals can perfectly reveal their actions and (ii) their records consist of the entire sequence of their unerased non-null signals, it extends to the case where each player's record is the \textit{summary statistics} of their unerased non-null signals. Theorem \ref{Theorem3} is also robust when both the signal $s_i$ and the record transition rule $\Psi_i$ are noisy, as long as both noises are sufficiently small and $s_i$ is drawn from a finite set.\footnote{Perfect monitoring and almost perfect monitoring are also the settings that
Clark, Fudenberg and Wolitzky (2021) focus on when they show the following positive result, that when records are first order,
players can sustain cooperation in strict equilibria in the strictly supermodular prisoner's dilemma.}   Nevertheless, under general imperfect monitoring (e.g., when the monitoring noise is large), constructing non-trivial equilibria in repeated games is technically challenging and whether players can sustain cooperation in these general settings remains an open question.\footnote{To the best of my knowledge, there is no folk theorem in community enforcement models with a continuum of players and general imperfect monitoring when the solution concept is \textit{steady state equilibrium}. This is because the algorithms in Abreu, Pearce and Stacchetti (1990) and Fudenberg, Levine and Maskin (1994) do not apply when there is an extra steady state condition on the record distribution, and constructing equilibria in repeated games with imperfect monitoring is not tractable in general.}

In what follows, I use the prisoner's dilemma as an example to
explain the intuition behind Theorems \ref{Theorem2} and \ref{Theorem3} as well as their comparisons with Theorem \ref{Theorem1}. For illustration purposes, I focus on the simple case where $\widehat{\delta}_1=\widehat{\delta}_2 \equiv \widehat{\delta}$, $\overline{\delta}_1=\overline{\delta}_2 \equiv \overline{\delta}$, namely, both populations share the same time preference and survival probability, and each player's record consists of the sequence of his unerased non-null signals.

First, Theorem \ref{Theorem3} suggests that communities
consisting of members with \textit{intermediate} expected lifespans can sustain cooperation even when people can selectively disclose their past actions and only \textit{first-order records} are available. This conclusion stands in contrast to Theorem \ref{Theorem1} that the maximal level of cooperation vanishes to $0$ as players become sufficiently long-lived, regardless of their record systems.

The comparison between Theorems \ref{Theorem1} and \ref{Theorem3} suggests that the maximal level of cooperation a community can sustain is \textit{not monotone} with respect to the expected lifespans of its members. In particular, fixing players' stage-game payoffs and time preference $\widehat{\delta} \in (0,1)$, the maximal probability with which players play $C$ in equilibrium is strictly greater than $0$ when $\overline{\delta}$ is intermediate but converges to $0$ as $\overline{\delta}$ goes to $1$.

Such  non-monotonicity stands in contrast to the standard conclusions in the theories of repeated games and community enforcement that players have stronger incentives to cooperate when they have higher effective discount factors.
The intuition is that while players with very short expected lifespans (e.g., those with $\overline{\delta}_i$ close to $0$) have no incentive to cooperate due to their impatience, players with sufficiently long expected lifespans also cannot sustain cooperation since (i) players with long good records will receive high continuation values and they will have no incentive to cooperate when they can sustain their current records by not disclosing new signals and (ii) when players are sufficiently long-lived, either they take cooperative actions with low probabilities when they do not have high continuation values or there will be a significant mass of old players with long good records and high continuation values who have no incentive to cooperate.

Second, the comparison between Theorems \ref{Theorem2} and \ref{Theorem3} suggests that players can sustain cooperation in purifiable equilibria in the submodular prisoner's dilemma but cannot sustain any cooperation in the supermodular prisoner's dilemma.\footnote{For example in the supermodular prisoner's dilemma, one can construct equilibria in which the average probability with which players play $C$ is strictly positive as long as $g_i \leq l_i \leq 1+g_i$  for every $i \in I$ although those equilibria are not purifiable.}
This stands in contrast to the takeaways from Takahashi (2010), Heller and Mohlin (2018), and Clark, Fudenberg and Wolitzky (2021) that when records are first order, it is easier to sustain cooperation when payoffs are supermodular compared to the case where payoffs are submodular. For example, Takahashi (2010) and Clark, Fudenberg and Wolitzky (2021) show that in the prisoner's dilemma game where records are \textit{first order} and players \textit{cannot} erase signals, there exist grim trigger equilibria and strict equilibria, respectively, that can sustain a positive level of cooperation \textit{if and only if} players' payoffs are strictly supermodular.\footnote{In Clark, Fudenberg and Wolitzky (2021), there exist purifiable equilibria where players can sustain some cooperation in the submodular prisoner's dilemma, such as the one constructed in my proof of Theorem \ref{Theorem3}, although none of these equilibria is strict.}
Heller and Mohlin (2018) study a community enforcement model in which with positive probability, each player is one of the several
 \textit{mixed-strategy commitment types} and each player can observe $k$ random samples of their partner's past play before choosing their actions (i.e., players' records are first order). They show that players can sustain some cooperation in the supermodular prisoner's dilemma but cannot sustain any cooperation in the submodular prisoner's dilemma.

Comparing my Theorems \ref{Theorem2} and \ref{Theorem3} to the results in Clark, Fudenberg and Wolitzky (2021) on first-order records, their model does not allow for endogenous record disclosure and their results focus on strict equilibria, a more demanding solution concept. In contrast, the players in my model can erase signals from their records, which imposes a monotonicity constraint on their continuation values. I show that (i) all cooperative equilibria in the supermodular prisoner's dilemma are \textit{not} robust when players can endogenously decide whether to disclose their signals and (ii) there exist cooperative equilibria in the submodular's prisoner's dilemma that are purifiable although players do not always have strict incentives.\footnote{One can show that in the prisoner's dilemma where players have regular record systems and can replace their signals with the null signal, they will always defect in \textit{all} strict equilibria. This together with Theorem \ref{Theorem2} implies that players cannot sustain any cooperation in \textit{any} strict equilibrium in any prisoner's dilemma. This observation provides a justification for focusing on purifiable equilibria in my analysis of Section \ref{sec4}.}

For some intuition behind this comparison, recall from the proof of Theorem \ref{Theorem1} that player $1$ has a strict incentive to defect when his record system is regular and his continuation value reaches its maximum. In order to deliver a high continuation value to players in population $1$ who are supposed to receive their highest continuation value, there must exist some players from population $2$ who will cooperate with them with strictly higher probability than with players in population $1$ who are supposed to receive the lowest continuation value. When actions are complements, any player who is willing to cooperate with an opponent who always defects must also have an incentive to cooperate with any opponent with any record.

I use a property of purifiable equilibria noted by Bhaskar, Mailath and Morris (2013) and Bhaskar and Thomas (2019) that when an equilibrium in a dynamic game is purifiable, players' actions cannot depend on payoff-irrelevant histories.
In my community enforcement game, an implication of purifiability is that when player $2$'s record is \textit{first order}, an equilibrium is purifiable only when for every $r_1,r_1' \in R_1$  and $r_2 \in R_2$, if player $1$'s expected actions are the same under $(r_1,r_2)$ and under $(r_1',r_2)$, then player $2$'s (potentially mixed) actions must be the same  under $(r_1,r_2)$ and under $(r_1',r_2)$. When players' actions are complements, I show in Lemma \ref{L1} that when player $1$ cooperates with weakly higher probability at
$(r_1,r_2)$ compared to $(r_1',r_2)$, player $2$ must also cooperate with weakly higher probability at
$(r_1,r_2)$ compared to $(r_1',r_2)$.


The above reasoning suggests that the player who is supposed to receive the highest continuation value will receive a stage-game payoff that is weakly less than that of any player from the same population with any record. Since continuation values are non-decreasing over time,\footnote{One caveat is that when the record transition rule is stochastic, although a player's continuation value must be non-decreasing \textit{in expectation}, it may strictly decrease following some record that occurs with positive probability. As a result, the player who is supposed to receive the highest continuation value may not obtain weakly lower stage-game payoff \textit{at every subsequent history} compared to a player with a generic record. I show that under the hypothesis that there exists an equilibrium with cooperation, for every record $r_i$ where player $i$'s continuation value approaches its maximum, there exists another record $\widehat{r}_i$ under which player $i$'s continuation value is significantly greater. This contradicts the requirement that each player's continuation value is bounded.} his \textit{discounted average payoff} is also weakly lower than the continuation value of any other player from the same population.
This contradicts the fact that in any equilibrium where players from a given population take cooperative actions with strictly positive probability, the highest continuation value in that population must be \textit{strictly} greater than the lowest.

When players' actions are strategic substitutes, consider the following strategy profile where both player roles cooperate with positive probability. Players in each population are divided into two subgroups: \textit{juniors} with no $C$ in their records and \textit{seniors} with at least one $C$ in their records. Seniors play $D$ against all opponents. Juniors play $C$ against seniors and mix between $C$ and $D$ against other juniors. Due to strategic substitutability, when a junior is indifferent between $C$ and $D$ against another junior who mixes between $C$ and $D$, he will have a strict incentive to play $C$ against seniors who always play $D$.

In order for the above strategy profile to be part of an equilibrium, players' expected lifespans cannot be too short. This is because otherwise, their effective discount factors will be too low, which precludes their incentives to cooperate. Their expected lifespans cannot be too long since otherwise, there will be too many seniors in the population who are unwilling to cooperate. Anticipating this, the juniors will also have no incentive to cooperate since their continuation value once they become senior is too low.
When players' expected lifespans are intermediate, this strategy profile is an equilibrium and such
an equilibrium is purifiable given that all incentives are strict except for the case in which two juniors are matched.

Note that in the cooperative equilibria I constructed for the strictly submodular prisoner's dilemma,
players do \textit{not} need to erase any signal on the equilibrium path. Moreover, those equilibria are also equilibria in the game where players cannot erase signals. However, these facts do not imply that players' ability to erase records is not relevant or not economically meaningful: In fact, they lead to novel constraints on what can be achieved in equilibrium and these constraints rule out, for example, the usual type of cooperative equilibria such as those where players use grim-trigger strategies.

Since Theorem \ref{Theorem3} suggests that players can sustain a positive level of cooperation when their payoffs are strictly monotone-submodular, one may wonder whether there is a folk theorem in the sense that players can cooperate with probability close to $1$, at least under some time preferences and survival probabilities.

Focusing on the prisoner's dilemma, I show that the answer to the above question is negative: As long as there is \textit{one} population of players who can erase their signals and have regular record systems, there exists a \textit{uniform} upper bound $q \in (0,1)$ on \textit{both} populations' probabilities of cooperation. This upper bound is also uniformly applicable to \textit{all} steady state equilibria under \textit{all} time preferences, expected lifespans, technologies that monitor their behaviors, and record transition rules.
\begin{Theorem}\label{Theorem4}
Fix any $(g_1,g_2,l_1,l_2) \in \mathbb{R}_{++}^4$. There exists $q \in (0,1)$ such that for any signal distribution $(f_1,f_2)$, any record systems such that player $1$'s record system is regular, any time preference and survival probability for both players, and any steady state equilibrium under these parameters, the average probability with which player $i$ plays $C$ is no more than $q$ for every $i \in \{1,2\}$.
\end{Theorem}
The proof is in Appendix \ref{secD}. This result is not implied by Theorem \ref{Theorem1} since Theorem \ref{Theorem1} requires
that
$\overline{\delta}_i$ converging to $1$ at a faster rate than $\widehat{\delta}_i$ converging to $1$, while the upper bound in Theorem \ref{Theorem4} applies \textit{uniformly} to all values of $(\overline{\delta}_i,\widehat{\delta}_i) \in [0,1)^2$.

The intuition for why player $1$'s probability of cooperation must be bounded below $1$ is similar to the intuition behind Theorem \ref{Theorem1}, that player $1$'s continuation value is non-decreasing over time and there exist at most a bounded number of periods that player $1$ has an incentive to cooperate.

As for player $2$'s probability of cooperation, suppose by way of contradiction that there exists an equilibrium in which player $2$ cooperates with an average probability that is close to $1$.
Then with probability close to $1$, player $1$ will have records that will induce player $2$ to take the cooperative action with probability close to $1$. When player $1$'s record system is regular, he can sustain his current continuation value by defecting in every subsequent period and then erasing his signal, from which he can secure a payoff that is arbitrarily close to $1+g_1$. In equilibrium, player $1$ can attain this payoff only when the average probability with which he plays $D$ is close to $1$. Anticipating this, player $2$ will have no incentive to cooperate with player $1$, which contradicts the hypothesis that player $2$ will cooperate with probability close to $1$.

Theorem \ref{Theorem4} provides a uniform upper bound on the frequency of cooperation when at least one population of players have regular record systems. A natural follow-up question is: What is the maximal equilibrium frequency of cooperation? Unfortunately, answering this question is beyond the scope of this paper since very little is known about the structure of steady state equilibria. For example, even when players cannot erase signals and their signals are publicly observed, it is unclear how to characterize the set of steady state equilibrium payoffs even in the limit as the effective discount factor goes to $1$. Players' ability to erase signals introduces a monotonicity constraint, which makes characterizing equilibrium payoffs even harder.

\section{Concluding Remarks}\label{sec5}
I analyze community enforcement models with a continuum of players where players endogenously decide whether to disclose signals about their past actions.
When a population of players can sustain their current records by not disclosing newly generated signals,
their continuation values must be \textit{non-decreasing} over time.  As a result, the only way to provide those players intertemporal incentives is by promising them higher continuation values in the future. This rules out their incentives to cooperate when they reach their highest continuation values and implies that communities with sufficiently long-lived players will almost never cooperate. In contrast, players with \textit{intermediate} expected lifespans can sustain a positive level of cooperation. Furthermore, there exist cooperative equilibria that are robust to a small amount of private payoff information when players' actions are substitutes but not when their actions are complements.

My conclusions stand in contrast to the takeaways from most of the existing works on repeated games and community enforcement,
that a higher \textit{effective discount factor} cannot decrease the maximal level of cooperation, no matter whether it results from a higher survival probability or a higher time preference, and that the complementary of players' actions can facilitate communities' abilities to sustain cooperation. I conclude with a discussion on the robustness of my results to alternative modeling assumptions.

\paragraph{Alternative Models of Disclosure:} My results extend to several alternative models of record disclosure. These include (i) players cannot erase signals but in every period after being matched, each player decides which subset of his past signals to disclose to their current partners, which is an analogue of the static disclosure games studied by Dzuida (2011) and Gao (2024) and (ii) players face constraints when disclosing their past signals, for example, each player can disclose at most $K$ signals to his current partners, which is the case studied by Di Tillio, Ottaviani and S{\o}rensen (2021).

The reason is that a key property of my baseline model is preserved in these alternative models, that each player's continuation value must be \textit{non-decreasing} over time. As a result, each player can ensure his current continuation value in the next period regardless of the action he takes in the current period.

\paragraph{Regularity Condition on Records:} My regularity condition on record systems requires that a player's record remains
unchanged if he generates the null signal, and in my model, he can replace any of his signal with the null signal. This condition rules out records that reveal \textit{the number of null signals} players have generated (e.g., the number of times that consumers do not post reviews) as well as records that reveal the \textit{exact date} at which each signal arrives (e.g., the exact date at which each review was posted). In these situations, players can no longer sustain their current records regardless of the actions they take, since they can be penalized, for example, when they generate the null signal or when all of their non-null signals are too old.
If this is the case, then their expected continuation values may \textit{decrease} over time.

Hence, my model fits when information about the timing of other players' signals is \textit{either} not available (such as eBay that only discloses the number of positive and negative reviews each seller has received) \textit{or} when such information is available online but it is very time-consuming for players to process such information (such as Yelp that discloses the exact date of each review, but it is very costly for most consumers to learn the exact timing of each review when the number of reviews is large). In these settings,
my result implies that it is beneficial for online platforms to delete old reviews, such as only disclosing the average rating in the last $6$ months.\footnote{In practice, platforms may not want to forbid consumers from deleting the reviews they posted due to other concerns, such as some of the reviews are defamatory, slanderous, or are illegal in other ways. Allowing the consumers to erase their reviews gives the sellers opportunities to bribe consumers for erasing negative reviews, which motivates the analysis in my paper.} In this respect, my model suggests a novel rationale for limited memories, which stands in contrast to the ones in the existing literature such as the seller's type is changing over time. In contrast, committing to disclose only the last $K$ reviews cannot help since the seller can still sustain his current record regardless of the action he takes.

\paragraph{Time Preference \textit{vs} Survival Probability:} In contrast to standard repeated game models, the equilibrium outcomes of my model depend not only on players' effective discount factor but also on whether it results from their time preferences or their survival probabilities. For example, when player $i$'s survival probability $\overline{\delta}_i$ goes to $1$ and his time preference $\widehat{\delta}_i$ is bounded below $1$, Theorem \ref{Theorem1} suggests that he will almost always take his strictly dominant action. In contrast, when player $i$'s survival probability belongs to some intermediate range and his time preference goes to $1$, my proof of Theorem \ref{Theorem3} suggests that there exists a purifiable equilibrium in which the probability of cooperation is bounded above $0$.

One natural question is: what will happen in the case where $\widehat{\delta}_i=1$ (i.e., player $i$ is infinitely patient) and the case where $\overline{\delta}_i=1$ (i.e., player $i$ is infinitely long-lived)?
When $\overline{\delta}_i=1$, the steady state distribution over records is mathematically not well-defined, although my negative results Theorems \ref{Theorem1}, \ref{Theorem2}, and \ref{Theorem4} apply to any $\overline{\delta}_i$ that is arbitrarily close to $1$. Theorem \ref{Theorem3} requires players' expected lifespans to be intermediate, which precludes the case where $\overline{\delta}_i$ is arbitrarily close to $1$.
When $\widehat{\delta}_i=1$ and $\overline{\delta}_i$ is bounded away from $1$, Theorem \ref{Theorem1} fails since it holds \textit{if and only if} $\frac{1-\overline{\delta}_i}{1-\widehat{\delta}_i}$ goes to $0$, i.e., $\overline{\delta}_i$ goes to $1$ at a faster rate compared to $\widehat{\delta}_i$ goes to $1$. Theorems \ref{Theorem2}, \ref{Theorem3}, and \ref{Theorem4} remain valid since none of their proofs relies on the assumption that $\widehat{\delta}_i<1$.


\paragraph{Comparison with Models \textit{without} Endogenous Disclosure:} Compared to Clark, Fudenberg and Wolitzky (2021), the players in my model can choose whether to disclose their signals. The cooperative purifiable equilibria constructed in the proof of Theorem \ref{Theorem3} remain equilibria in their model, although all the cooperative strict equilibria in their model are not robust when there is endogenous record disclosure.

One may wonder whether the set of steady state equilibria in Clark, Fudenberg and Wolitzky (2021) is a superset of that in my model given that any equilibrium in my model must satisfy an additional monotonicity constraint that each player's continuation value must be non-decreasing over time. The answer to that question is \textit{no}. To see this, take the leading example in which each player's signal coincides with his action and his record consists of the sequence of his actions.
By observing the entire sequence of a player's actions as in Clark, Fudenberg and Wolitzky (2021), one may not perfectly learn the sequence of his \textit{unerased} actions since that sequence is also affected by the decisions to erase signals and in equilibrium, players may randomize when deciding whether to erase each signal. As an example, consider the following stage game and the following equilibrium in my model:
\begin{center}
\begin{tabular}{| c | c | c |}
  \hline
 - & C & D \\
  \hline
 A & $1, 0$ & $-1, 0$ \\
  \hline
  B & $-1, 0$ & $1,0$ \\
  \hline
\end{tabular}
\end{center}
Each player 2 plays $C$ in the first period of his life and in the future, his action depends only on his own record in which he plays $C$ when his record is the empty set and plays $D$ otherwise. Player $2$ erases $D$ for sure at every record, erases $C$ for sure at every record that is non-empty, and erases $C$ at the empty record with probability $p \in (0,1)$, where $p$ is chosen such that the unconditional probability that player $1$ having an empty record is $1/2$. Player $1$ plays $A$ when player $2$'s record is the empty set and plays $B$ otherwise.

When player $1$ can only observe the sequence of player $2$'s actions but not player $2$'s erasing decisions, the above strategy profile is no longer feasible since player $1$ cannot perfectly infer what player $2$'s record is in my model by observing only the sequence of player $2$'s unerased actions.

\appendix
\section{Proof of Corollary 1}\label{secA}
 Recall that in any \textit{finite} normal-form game, a pure action $a_i \in A_i$ is strictly dominated in the stage game \textit{if and only if} it is never a best reply and there exists $\eta>0$ such that regardless of player $-i$'s action $\alpha_{-i} \in \Delta (A_{-i})$, player $i$'s payoff from playing $a_i$ is less than his payoff from playing a best reply minus $\eta$.

 Using the same argument as that in the proof of Theorem \ref{Theorem1}, one can show that for every $\widehat{\delta}_i \in (0,1)$, there exists $\delta^* \in (0,1)$ such that when $\overline{\delta}_i  > \delta^*$, the average probability with which player $i$ takes strictly dominated actions is less than $\varepsilon$ in all equilibria. Let $A_i^1 \subset A_i$ denote the set of player $i$'s actions that survive the first round of deletion but does not survive the second round. If $A_i^1$ is non-empty, then there exists $\eta>0$ that depends only on $u_i$ such that all actions in $A_i^1$ are still strictly dominated by at least $\eta$ when the probability that player $-i$ plays strictly dominated actions is no more than $\eta$. According to the Markov's inequality,
if the average probability with which player $-i$ plays strictly dominated actions is no more than $\varepsilon$, then
histories where player $-i$ plays strictly dominated actions with probability more than $\eta$ occurs with probability less than $\varepsilon/\eta$. Using the argument in Theorem \ref{Theorem1}, we know that
for every $\widehat{\delta}_i \in (0,1)$, there exists $\delta^* \in (0,1)$ such that when $\overline{\delta}_i  > \delta^*$,
the probability that player $i$ takes actions in $A_i^1$ is at most $\varepsilon + \varepsilon/\eta$.
Iterate this process, since every non-rationalizable action in the stage game can survive at most $2n$ rounds of iterative deletion of strictly dominated strategies with
$n \equiv \max \{|A_1|,|A_2|\}$, we know that for every $\widehat{\delta}_1, \widehat{\delta}_2 \in (0,1)$, there exists $\delta^* \in (0,1)$ such that when
 $\overline{\delta}_1,\overline{\delta}_2> \delta^*$,
 the probability with which player $i$ takes non-rationalizable actions is bounded above by a linear function of $\varepsilon$, with the linear coefficient depending only on $(u_1,u_2,\widehat{\delta}_1,\widehat{\delta}_2)$. This leads to the conclusion of Corollary \ref{Cor1}.

\section{Proof of Theorem 2}\label{secB}
Suppose by way of contradiction that there exists a purifiable equilibrium $(\sigma_1,\sigma_2,\mu)$ in which player $1$ plays $a_1 \neq a_1^*$ with strictly positive probability at some record profile $(r_1,r_2)$ that satisfies $\mu(r_1,r_2)>0$.
Let $V_1(r_1)$ denote player $1$'s continuation value when his record is $r_1$, before knowing whether he will be matched and what his partner's record is.
Let $\overline{V}_1 \equiv \sup_{r_1 \in R_1} V_1(r_1)$ and let $\underline{V}_1 \equiv \inf_{r_1 \in R_1} V_1(r_1)$. Player $1$'s incentive to play $a_1 \neq a_1^*$ at some $(r_1,r_2)$ implies that $\overline{V}_1>\underline{V}_1$. Let
\begin{equation*}
c^* \equiv \min_{a_1 \neq a_1^*, a_{2} \in A_{2}} \Big\{
u_1 (a_1^*,a_{2})- u_1 (a_1,a_{2})
\Big\}.
\end{equation*}
Since player $1$'s payoff is monotone, $a_1^*$ is player $1$'s strictly dominant action, and the stage game is finite, the value of $c^*$ must be strictly positive. Fix any $\varepsilon$ that satisfies
\begin{equation}\label{4.1}
0<\varepsilon < \min \Big\{\frac{\overline{V}_1-\underline{V}_1}{3}, \frac{1-\delta_1}{2\delta_1} c^*\Big\}.
\end{equation}
By definition, there exist $\overline{r}_1 \in R_1$ with $V_1(\overline{r}_1) \in [\overline{V}_1-\varepsilon,\overline{V}_1]$ as well as $\underline{r}_1 \in R_1$ with $V_1(\underline{r}_1) \in [\underline{V}_1,\underline{V}_1+\varepsilon]$. My earlier requirement on
$\varepsilon$ in (\ref{4.1}) implies that $V_1(\overline{r}_1)>V_1(\underline{r}_1)$.
When player $1$'s record system is regular,
he will not take any action  $a_1 \neq a_1^*$ with positive probability at record $\overline{r}_1$
regardless of the record of player $2$ he is matched with.
This is because player $1$'s continuation value from playing $a_1$
is at most
\begin{equation}\label{4.2}
(1-\delta_1) u_1(a_1,\sigma_2^a(\overline{r}_1,r_2)) + \delta_1 \overline{V}_1,
\end{equation}
where $\sigma_2^a(\overline{r}_1,r_2) \in \Delta (A_2)$ is player $2$'s action at $(\overline{r}_1,r_2)$. In contrast, player $1$'s continuation value from playing $a_1^*$ and erasing every signal he generates is at least $(1-\delta_1) u_1(a_1^*,\sigma_2^a(\overline{r}_1,r_2)) + \delta_1 (\overline{V}_1-\varepsilon)$, which is strictly greater than (\ref{4.2}).
The following lemma is the only step where I use the purifiability refinement:
\begin{Lemma}\label{L1}
Fix any purifiable equilibrium. For every $\overline{r}_1, \underline{r}_1 \in R_1$ that satisfy   $V_1(\overline{r}_1) \geq \overline{V}_1-\varepsilon$ and $V_1(\underline{r}_1) <\underline{V}_1+\varepsilon$ as well as every $r_2 \in R_2$, player $2$'s action at $(\underline{r}_1,r_2)$ is weakly greater than his action at $(\overline{r}_1,r_2)$ in the sense of FOSD.
\end{Lemma}
\begin{proof}
Since player $1$'s payoff is monotone and he strictly prefers to take action $a_1^*$ at $\overline{r}_1$, we know that for every $r_2$ that occurs with positive probability,
player $1$'s action at $(\underline{r}_1,r_2)$, denoted by $\alpha_1 \in \Delta (A_1)$ weakly FOSDs his action at $(\overline{r}_1,r_2)$.
Player $2$'s discounted average payoff from playing $a_2$ at  $(\overline{r}_1,r_2)$ is
\begin{equation}\label{4.3}
(1-\delta_2) u_2(a_1^*,a_2) + \delta_2 \mathbb{E}[V_2|a_1^*,\overline{r}_1,a_2,r_2],
\end{equation}
where $\mathbb{E}[V_2|a_1,r_1,a_2,r_2]$ stands for player $2$'s continuation value in the next period conditional on players' current-period actions and records, taking into account player $2$'s decision to erase his signals optimally.
His payoff from playing $a_2$ at $(\underline{r}_1,r_2)$ is
\begin{equation}\label{4.4}
(1-\delta_2) u_2(\alpha_1,a_2) + \delta_2 \mathbb{E}[V_2|\alpha_1,\underline{r}_1,a_2,r_2].
\end{equation}
When player $2$'s records are first order, $\mathbb{E}[V_2|a_1,r_1,a_2,r_2]$ depends only on $(a_2,r_2)$, which I rewrite as $\mathbb{E}[V_2|a_2,r_2]$. Since $u_2(a_1,a_2)$ has weakly increasing differences, the following optimization problem
\begin{equation}
\max_{a_2 \in A_2} \Big\{ (1-\delta_2) u_2(a_1,a_2) + \delta_2 \mathbb{E}[V_2|a_2,r_2] \Big\}
\end{equation}
satisfies the single-crossing condition with respect to $a_1$. Since $\alpha_1 \succsim_1 a_1^*$,
the set of maximizers when $a_1=\alpha_1$, denoted by $A_2^{**}$, dominates
the set of maximizers when $a_1=a_1^*$, denoted by $A_2^*$, in strong set order (Milgrom and Shannon 1994). Consider any $\varepsilon$-perturbed stage game where player $2$'s stage-game payoff from playing $a_2$ is $u_2(a_1,a_2)+ \varepsilon z_2(a_2)$, where $z_2(a_2)$ has bounded support and a continuous distribution. Without loss of generality, suppose $z_2$ is always non-negative.
Fix any small enough $\varepsilon>0$ that satisfies:
\begin{equation*}
2 \varepsilon \max_{a_2 \in A_2} |z_2(a_2)| <
\max_{a_2 \in A_2} \Big\{ (1-\delta_2) u_2(a_1^*,a_2) + \delta_2 \mathbb{E}[V_2|a_2,r_2] \Big\}
-\max_{a_2 \in A_2\backslash A_2^*} \Big\{ (1-\delta_2) u_2(a_1^*,a_2) + \delta_2 \mathbb{E}[V_2|a_2,r_2] \Big\},
\end{equation*}
\begin{equation*}
2 \varepsilon \max_{a_2 \in A_2} |z_2(a_2)| <
\max_{a_2 \in A_2} \Big\{ (1-\delta_2) u_2(\alpha_1,a_2) + \delta_2 \mathbb{E}[V_2|a_2,r_2] \Big\}
-\max_{a_2 \in A_2\backslash A_2^{**}} \Big\{ (1-\delta_2) u_2(\alpha_1,a_2) + \delta_2 \mathbb{E}[V_2|a_2,r_2] \Big\},
\end{equation*}
which implies that in the repeated $\varepsilon$-perturbed game, only actions in $A_2^*$ will be played with positive probability at $(\overline{r}_1,r_2)$ and only actions in $A_2^{**}$ will be played with positive probability at $(\underline{r}_1,r_2)$.
Fix any $a_2 \in A_2^*$. The probability that player $2$'s action at $(\overline{r}_1,r_2)$ being weakly greater than $a_2$ equals the probability of
\begin{equation}\label{low}
\max_{\widetilde{a}_2 \succsim a_2, \widetilde{a}_2 \in A_2^*} z_2 (\widetilde{a}_2)
\geq \max_{\widetilde{a}_2 \prec a_2, \widetilde{a}_2 \in A_2^*} z_2 (\widetilde{a}_2).
\end{equation}
The probability that player $2$'s action at $(\underline{r}_1,r_2)$ being weakly greater than $a_2$ equals the probability of
\begin{equation}\label{high}
\max_{\widetilde{a}_2 \succsim a_2, \widetilde{a}_2 \in A_2^{**}} z_2 (\widetilde{a}_2)
\geq \max_{\widetilde{a}_2 \prec a_2, \widetilde{a}_2 \in A_2^{**}} z_2 (\widetilde{a}_2).
\end{equation}
Let $\widehat{a}_2$ denote the highest element of $A_2^*$ that is strictly below $a_2$. I consider three cases separately.
If $a_2 \notin A_2^{**}$, then  given that $A_2^{**}$ dominates $A_2^*$ in strong set order, no element that is weakly below $a_2$ belongs to $A_2^{**}$, in which case the probability of (\ref{high}) equals $1$.
If  $\widehat{a}_2 \notin A_2^{**}$, then no element that is strictly below $a_2$ belongs to $A_2^{**}$, in which case the probability of (\ref{high}) also equals $1$. In these two cases, the probability of (\ref{high}) is weakly greater than the probability of (\ref{low}).
If $a_2,\widehat{a}_2 \in A_2^{**}$, then for every $a_2' \succ a_2$ that satisfies $a_2' \in A_2^*$, it must also be the case that $a_2' \in A_2^{**}$, and for every $a_2'' \prec \widehat{a}_2$ that satisfies $a_2'' \in A_2^{**}$, it must also be the case that $a_2'' \in A_2^*$. This implies that the following event occurs with probability $1$:
\begin{equation}\label{4.8}
\max_{\widetilde{a}_2 \succsim a_2, \widetilde{a}_2 \in A_2^{**}} z_2 (\widetilde{a}_2) \geq \max_{\widetilde{a}_2 \succsim a_2, \widetilde{a}_2 \in A_2^*} z_2 (\widetilde{a}_2) \quad and \quad
\max_{\widetilde{a}_2 \prec a_2, \widetilde{a}_2 \in A_2^*} z_2 (\widetilde{a}_2) \geq \max_{\widetilde{a}_2 \prec a_2, \widetilde{a}_2 \in A_2^{**}} z_2 (\widetilde{a}_2).
\end{equation}
As a result, event (\ref{high}) occurs whenever event (\ref{low}) occurs, which implies that
the probability of event (\ref{high}) is weakly greater than the probability of event (\ref{low}). This further implies that player $2$'s mixed action at $(\underline{r}_1,r_2)$ weakly FOSDs his  mixed action at $(\overline{r}_1,r_2)$.
\end{proof}
Since player $1$ plays $a_1^*$ for sure when his record is $\overline{r}_1$,
his discounted average payoff at record $\overline{r}_1$ when he uses his equilibrium strategy is
\begin{equation}\label{4.7}
p \sum_{r_2 \in R_2} \mu_2(r_2) \Big\{
(1-\delta_1) u_1(a_1^*, \sigma_2^a(\overline{r}_1,r_2)) + \delta_1 \mathbb{E}[V_1|a_1^*,\overline{r}_1,\sigma_2^a(\overline{r}_1,r_2),r_2]
\Big\} + (1-p) \delta_1 V_1(\overline{r}_1),
\end{equation}
where $\mathbb{E}[V_1|a_1,r_1,a_2,r_2]$ stands for player $1$'s continuation value in the next period conditional on the record and action profiles in the current period, taking into account his decisions to erase his signals optimally.
If player $1$'s current-period record is $\underline{r}_1$ and he deviates by playing $a_1^*$ and erasing every signal he generates,
then his record remains $\underline{r}_1$ given that his record system is regular. Therefore,
his discounted average payoff at $\underline{r}_1$ under such a deviation is
\begin{equation}\label{4.8}
p \sum_{r_2 \in R_2} \mu_2(r_2) \Big\{
(1-\delta_1) u_1(a_1^*, \sigma_2^a(\underline{r}_1,r_2)) + \delta_1 V(\underline{r}_1)
\Big\} + (1-p) \delta_1 V_1(\underline{r}_1).
\end{equation}
By definition, the value of (\ref{4.8}) is weakly less than player $1$'s continuation value at record $\underline{r}_1$, which I denote by $V_1(\underline{r}_1)$. According to Lemma \ref{L1}, $\sigma_2^a(\underline{r}_1,r_2)$ weakly FOSDs $\sigma_2^a(\overline{r}_1,r_2)$.
This together with
the assumption that $u_1$ is monotone implies that
$ u_1(a_1^*, \sigma_2^a(\overline{r}_1,r_2)) \leq u_1(a_1^*, \sigma_2^a(\underline{r}_1,r_2))$. By definition, $V_1(\overline{r}_1)>V_1(\underline{r}_1)$. Since the difference between (\ref{4.7}) and (\ref{4.8}) is at least $V_1(\overline{r}_1)-V_1(\underline{r}_1)$, one can obtain the following inequality by subtracting (\ref{4.8}) from (\ref{4.7}):
\begin{equation}\label{4.5}
p \mathbb{E}[V_1|a_1^*,\overline{r}_1]+ (1-p)V_1(\overline{r}_1)- V_1(\underline{r}_1) \geq \frac{V_1(\overline{r}_1)-V_1(\underline{r}_1)}{\delta_1},
\end{equation}
where
\begin{equation*}
\mathbb{E}[V_1|a_1^*,\overline{r}_1] \equiv \sum_{r_2 \in R_2} \mu_2(r_2)\mathbb{E}[V_1|a_1^*,\overline{r}_1,\sigma_2^a(\overline{r}_1,r_2),r_2].
\end{equation*}
Let $R_1(\overline{r}_1)$ denote the set of player $1$'s records that occur with positive probability in the next period
when his current-period record is $\overline{r}_1$ and he plays his equilibrium strategy. Inequality (\ref{4.5}) implies that
\begin{equation}\label{4.6}
\max_{r_1 \in R_1(\overline{r}_1)} V_1(r_1)-V(\underline{r}_1) \geq
p \mathbb{E}[V_1|a_1^*,\overline{r}_1]+ (1-p)V_1(\overline{r}_1)- V_1(\underline{r}_1)
\geq \frac{V_1(\overline{r}_1)-V_1(\underline{r}_1)}{\delta_1}.
\end{equation}
Hence, for any $\overline{r}_1$ that satisfies $V_1(\overline{r}_1) \in [\overline{V}_1-\varepsilon,\overline{V}_1]$, there exists  $r_1' \in R_1(\overline{r}_1)$ such that $V_1(r_1')-V_1(\underline{r}_1) \geq \delta_1^{-1} (V_1(\overline{r}_1)-V_1(\underline{r}_1))$. The definition of $\overline{V}_1$ implies that for every $\varepsilon>0$, there exists $r_1 \in R_1$ such that $V_1(r_1)> \overline{V}_1-\varepsilon$. Fix any $\underline{r}_1$ that satisfies $V_1(\underline{r}_1)> \underline{V}_1-\varepsilon$ and
pick $\overline{r}_1$ such that $\delta_1^{-1}( V_1(\overline{r}_1)- V_1(\underline{r}_1)) > \overline{V}_1-V_1(\underline{r}_1)$, my earlier conclusion implies that there exists $r_1' \in R_1(\overline{r}_1)$ such that
$V_1(r_1')-V_1(\underline{r}_1) \geq \delta_1^{-1} (V_1(\overline{r}_1)-V_1(\underline{r}_1)) >\overline{V}_1-V_1(\underline{r}_1)$. This implies that $V_1(r_1')> \overline{V}_1$, which leads to a contradiction.

\section{Proof of Theorem 3}\label{secC}
Recall the definitions of $(a_1^*,a_2^*)$ and $(\overline{a}_1,\overline{a}_2)$. Without loss of generality, I normalize players' payoffs so that $u_i(a_1^*,a_2^*)=0$ and $u_i(\overline{a}_1,\overline{a}_2)=1$ for every $i \in I$. Let $g_i \equiv u_i(a_i^*,\overline{a}_2)-u_i(\overline{a}_i,\overline{a}_2)$ and let  $l_i \equiv u_i(a_i^*,a_2^*)-u_i(\overline{a}_i,a_2^*)$. Players' stage-game payoffs from action profiles $\{a_1^*,\overline{a}_1\} \times \{a_2^*,\overline{a}_2\}$ are
\begin{center}
\begin{tabular}{| c | c | c |}
\hline
 - & $\overline{a}_2$ & $a_2^*$  \\
  \hline
  $\overline{a}_1$ & $1,1$ & $-l_1,1+g_2$ \\
  \hline
  $a_1^*$ & $1+g_1,-l_2$ & $0,0$  \\
  \hline
\end{tabular}
\end{center}
Since $u_1$
and $u_2$
are both monotone and strictly submodular, we have $g_1>l_1>0$
and $g_2>l_2>0$.

For any $\widehat{\delta}_i \in (0,1)$ large enough, I construct a purifiable equilibrium under
an intermediate range of survival probabilities $[\delta',\delta''] \subset (0,1)$ such that (i) each player $i$ plays only $\overline{a}_i$ and $a_i^*$ with strictly positive probability and (ii) the average probability that player $i$ plays $\overline{a}_i$ is strictly positive. Players' strategies in this equilibrium is sequentially rational and moreover, the equilibrium is robust to small amount of noise in the signal structure, the record transition rule, and players' interpretations of others' signals.
For illustration purposes, my construction focuses on the case in which $p=1$, i.e., players are matched with probability $1$. Extensions to cases with $p \in (0,1)$ are straightforward.

Since each player's signal perfectly reveals his action,
I partition the set of records into two categories. I say that a player $i$ is \textit{junior} if there is no $\overline{a}_i$ in his record and is \textit{senior} otherwise. The number of other actions a player has in his record does \textit{not} affect whether he is senior or junior.
Senior player $i$ plays $a_i^*$ against all the partners that he is matched with. Junior player $i$ plays $\overline{a}_i$ against seniors in the other population and plays $\overline{a}_i$ with probability $q_i \in (0,1)$ and $a_i^*$ with complementary probability
against juniors in the other population. Let $\mu_i \in (0,1)$ denote the fraction of population $i$ who are juniors. Let $\overline{V}_i$ and $\underline{V}_i$ denote the continuation values of a senior player $i$ and a junior player $i$, respectively.

Under such a strategy profile, it is straightforward that any action other than $a_i^*$ and $\overline{a}_i$ is strictly dominated by $a_i^*$ since player $i$'s payoff is monotone,
and moreover, players have no incentive to erase $\overline{a}_i$ and their decisions to erase other actions are irrelevant.
Each junior player $i$ is indifferent between $a_i^*$ and $\overline{a}_i$ when facing another junior, which leads to the following indifference condition:
\begin{equation*}
(1-\delta_i) u_i \Big(\overline{a}_i,q_{-i} \overline{a}_{-i} + (1-q_{-i}) a_{-i}^* \Big) + \delta_i \overline{V}_i=
(1-\delta_i) u_i \Big(a_i^*,q_{-i} \overline{a}_{-i} + (1-q_{-i}) a_{-i}^* \Big) + \delta_i \underline{V}_i,
\end{equation*}
which reduces to
\begin{equation}\label{4.9}
\overline{V}_i-\underline{V}_i = \frac{1-\delta_i}{\delta_i} \Big(
q_{-i} g_i + (1-q_{-i}) l_i
\Big).
\end{equation}
This leads to a lower bound on the effective discount factor $\delta_i$, which in turn leads to a lower bound on $\overline{\delta}_i$.

A junior's incentive to play a higher action $\overline{a}_i$ against another junior implies that he has a \textit{strict} incentive to play $\overline{a}_i$ against any senior. As a result, each junior player $i$'s continuation value satisfies
\begin{equation*}
\underline{V}_i = \mu_{-i} \Big\{
(1-\delta_i) u_i \Big(\overline{a}_i,q_{-i} \overline{a}_{-i} + (1-q_{-i}) a_{-i}^* \Big)
+ \delta_i \overline{V}_i
\Big\} + (1-\mu_{-i}) \Big\{ (1-\delta_i) u_i(\overline{a}_i,a_{-i}^*) +\delta_i \overline{V}_i
\Big\},
\end{equation*}
which from (\ref{4.9}), is equivalent to
\begin{equation}\label{4.10}
\underline{V}_i = q_j \Big\{
\mu_{-i} + g_i - (1-\mu_{-i}) l_i
\Big\}.
\end{equation}
Since a senior player $i$'s continuation value satisfies
\begin{equation}\label{4.11}
\overline{V}_i= \mu_{-i} u_i(a_i^*,\overline{a}_{-i}) + (1-\mu_{-i}) u_i(a_i^*,a_{-i}^*)=\mu_{-i} (1+g_i),
\end{equation}
equations (\ref{4.9}) and (\ref{4.10}) together imply that
\begin{equation*}
\mu_{-i} (1+g_i)= q_{-i} \Big\{
\mu_{-i} + g_i - (1-\mu_{-i}) l_i
\Big\}+  \frac{1-\delta_i}{\delta_i} \Big(
q_{-i} g_i + (1-q_{-i}) l_i
\Big),
\end{equation*}
or equivalently,
\begin{equation}\label{4.12}
(1-q_{-i}) \mu_{-i} = \frac{1-\delta_i}{\delta_i} \cdot \frac{l_i}{1+g_i} + q_{-i} \frac{g_i-l_i}{1+g_i} \Big(\frac{1}{\delta_i} -\mu_{-i} \Big).
\end{equation}
Since $g_i>l_i$, the LHS of (\ref{4.12}) is strictly decreasing in $q_{-i}$ and equals $0$ when $q_{-i}=1$ and the RHS of (\ref{4.12}) is strictly increasing in $q_{-i}$ and is always strictly positive. This implies that there exists a solution to (\ref{4.12})  if and only if the LHS is no less than the RHS when $q_{-i}=0$, or equivalently,
\begin{equation}\label{4.13}
\mu_{-i} \geq \frac{1-\delta_i}{\delta_i} \cdot \frac{l_i}{1+g_i} \textrm{ for every } i \in I,
\end{equation}
where the steady state record distributions for the players $\mu_i$ and $\mu_{-i}$ satisfy
\begin{equation}\label{4.14}
\mu_i = (1-\overline{\delta}_i) + \overline{\delta}_i \mu_{i} \mu_{-i} (1-q_{i}) \textrm{ for every } i \in I.
\end{equation}
Equation (\ref{4.14}) implies that $\mu_i \geq 1-\overline{\delta}_i$ for every $i \in I$, and therefore, it is sufficient to show that there exists a non-empty interval $[\delta',\delta''] \subset (0,1)$ such that when $\overline{\delta}_1,\overline{\delta}_2 \in [\delta',\delta'']$, we have
\begin{equation}\label{4.15}
\mu_{-i} \geq 1-\overline{\delta}_{-i} \geq \frac{1-\delta_i}{\delta_i} \cdot \frac{l_i}{1+g_i}.
\end{equation}
This is indeed the case when $\widehat{\delta}_i$ are close enough to $1$ for every $i \in I$, in which case (\ref{4.15}) is satisfied as long as $\overline{\delta}_i$ is not too close to $1$. Such an equilibrium is purifiable since players have strict incentives except for one information set, which is when a junior is matched with another junior.

\section{Proof of Theorem 4}\label{secD}
First, I bound the probability that player $1$ plays $C$ from above. Let $p_k$ denote player $1$'s expected probability of playing $C$ in the $k$th period of his life (conditional on him being active for at least $k$ periods).
Since player $1$'s lowest stage-game cost of cooperation is $\min\{g_1,l_1\}$, his highest continuation value is no more than $1+g_1$, and his lowest continuation value is no less than $0$, inequality (\ref{3.4}) implies that
\begin{equation}\label{4.20}
\sum_{k=0}^{+\infty} p_k \leq \frac{1+g_1}{\min\{g_1,l_1\}} \cdot \frac{\delta_1}{1-\delta_1}.
\end{equation}
Let $X \equiv \frac{1+g_1}{\min\{g_1,l_1\}}$, which depends only on $u_1$.
By the law of total probabilities, player $1$'s expected probability of playing $C$ is $\sum_{k=0}^{+\infty} (1-\overline{\delta}_1) \overline{\delta}_1^k p_k$, and by inequality (\ref{4.20}) and the fact that $\overline{\delta}_1 \geq
\overline{\delta}_1 \cdot \widehat{\delta}_1 \equiv
\delta_1$, we have
\begin{equation}\label{4.21}
\sum_{k=0}^{+\infty} (1-\overline{\delta}_1) \overline{\delta}_1^k p_k
\leq 1-\overline{\delta}_1^{\frac{X\delta_1}{1-\delta_1}} \leq 1-\delta_1^{\frac{X\delta_1}{1-\delta_1}}
\end{equation}
Since $\frac{X\delta_1}{1-\delta_1} \log \delta_1$ is decreasing in $\delta_1 \in [0,1)$, we know that for every $\delta_1 \in [0,1)$,
\begin{equation}\label{4.22}
 \frac{X\delta_1}{1-\delta_1} \log \delta_1 \geq \lim_{\delta_1 \rightarrow 1} \Big\{ \frac{X\delta_1}{1-\delta_1} \log \delta_1 \Big\} =-X,
\end{equation}
where the last equation is derived from L'Hospital Rule.
This implies that the RHS of (\ref{4.21}) is no more than $1-e^{-X}$, which leads to an upper bound on player $1$'s average probability of cooperation.

Next, I bound the probability that player $2$ plays $C$ from above. Suppose by way of contradiction that for every $\varepsilon>0$, there exists a parameter configuration of the game as well as a steady state equilibrium under that configuration in which the average probability with which player $2$ plays $C$ is more than $1-\varepsilon$. Let $\mu_1$ denote the steady state distribution of player $1$'s records. For every $\eta>0$,
let $R_{1,\eta}$ denote the set of player $1$'s records such that the probability with which player $2$ cooperates with each of these records is at least $1-\eta$. The Markov's inequality implies that $\mu_{1} (R_{1,\eta}) \geq 1-\frac{\varepsilon}{\eta}$.
Since player $1$'s record system is regular, he can secure an expected continuation value of at least
\begin{equation}
(1+g_1) \Big( 1-\frac{\varepsilon}{\eta} \Big) (1-\eta)
\end{equation}
by playing $D$ in every period and erasing every signal he generates. The above expression converges to $1+g_1$ when both $\varepsilon$ and $\eta$ go to $1$ but $\varepsilon$ vanishes at a faster rate. In order to attain such an average payoff of close to $1+g_1$ in equilibrium, the equilibrium probability with which player $1$ plays $D$ must be close to $1$. If player $2$ plays $C$ with probability more than $1-\varepsilon$ in equilibrium, his equilibrium payoff is strictly less than $0$ given that player $1$ plays $D$ with probability close to $1$. This leads to a contradiction since player $2$ can secure payoff $0$ by playing $D$ in every period.

\end{spacing}

\end{document}